\documentclass{aa}  

\usepackage{graphicx}
\usepackage{txfonts}
\usepackage{amsmath}	
\usepackage{amssymb}	
\usepackage{siunitx}
\usepackage{todonotes}
\usepackage{ulem}
\usepackage{subcaption}
\usepackage[perpage]{footmisc}

\defcitealias{Eastman1996}{E96}
\defcitealias{Dessart2005a}{D05}

\newcommand*\dif{\mathop{}\!\mathrm{d}}
\newcommand{\snia}{SN~Ia}
\newcommand{\snias}{SNe~Ia}

\newcommand{\sniasx}{SNe~Iax}
\newcommand{\snii}{SN~II}
\newcommand{\sniis}{SNe~II}

\newcommand{\sniisp}{SNe~IIP}
\newcommand{\tardis}{\textsc{tardis}}
\usepackage[draft=True]{hyperref}
\usepackage{cleveref}
\crefname{equation}{Eq.}{Eqs.}
\crefname{figure}{Fig.}{Figs.}
\crefname{section}{Sect.}{Sects.}
\crefname{table}{Table}{Tables}
\crefname{chapter}{Chapter}{Chapters}
\bibpunct{(}{)}{;}{a}{}{,} 

\begin{document}

   \title{Spectral modeling of type II supernovae}

   \subtitle{I. Dilution factors}

   \author{C. Vogl
          \inst{1,4}
          \and
          S. A. Sim\inst{2}
          \and
          U. M. Noebauer\inst{1}
          \and
          W. E. Kerzendorf\inst{3}
          \and
          W. Hillebrandt\inst{1}
          }

   \institute{Max-Planck-Institut f\"ur Astrophysik, 
   Karl-Schwarzschild-Str. 1, D-85741 Garching, Germany\\
              \email{cvogl@mpa-garching.mpg.de}
         \and
Astrophysics Research Centre, School of Mathematics and 
Physics, Queen's University Belfast, Belfast BT7 1NN, UK\\
             \email{s.sim@qub.ac.uk}
          \and
European Southern Observatory, Karl-Schwarzschild-Str. 2, 
D-85741 Garching, Germany
\and 
Physik Department, Technische Universit{\"a}t M{\"u}nchen, 
James-Franck-Str. 1, D-85741 Garching, Germany
             }

   \date{Received 25 Juni 2018 / Accepted 5 November 2018}

  \abstract{
  We present substantial extensions to the Monte 
  Carlo radiative transfer code \textsc{tardis} to 
  perform spectral synthesis for type II 
  supernovae. By incorporating a non-LTE ionization
  and excitation treatment for hydrogen, a full 
  account of free-free and bound-free processes, a 
  self-consistent determination of the thermal 
  state and by improving the handling of 
  relativistic effects, the improved code version 
  includes the necessary physics to perform 
  spectral synthesis for type II supernovae to high 
  precision as required for the reliable inference of
  supernova properties. We demonstrate the capabilities of the 
  extended version of \textsc{tardis} by 
  calculating synthetic spectra for the 
  prototypical type II supernova SN1999em and by 
  deriving a new and independent set of dilution 
  factors for the expanding photosphere method. We have 
  investigated in detail the dependence of the dilution factors 
  on photospheric properties and, for the first 
  time, on changes in metallicity. We also compare 
  our results with two previously published sets of 
  dilution factors by \citet{Eastman1996} and by 
  \citet{Dessart2005a}, and discuss the potential 
  sources of the discrepancies between studies.}


   \maketitle
%

\section{Introduction}
In recent years the availability of spectral data 
for hydrogen-rich supernovae (Type II; SNe II) has 
increased dramatically. Measurements for hundreds 
of SNe~II are now publicly accessible \citep[see 
e.g.,][for recent data releases]
{Poznanski2009,DAndrea2010,Hicken2017}, providing a 
dataset that contains a wealth of information about 
the kinematics of the explosion, the progenitor 
systems \citep[e.g.,][]{Jerkstrand2012}, the 
circumstellar material \citep[e.g.,][]{Quimby2007}, 
and much more. Most of the analysis of this data 
has focused on the study of easily measurable 
spectral parameters such as line absorption 
velocities or equivalent widths and their 
correlations
\citep[see e.g.,][]{Gutierrez2017a,Gutierrez2017b}, 
omitting the full information contained in the 
spectra. To establish connections between these 
parameters and the underlying quantities, such as 
the metallicity, most studies rely on approximate 
relations that have been calibrated based on 
theoretical models \citep[see e.g.,][]{Anderson2016}.
In contrast, only a few well-observed type II 
supernovae, such as SN1999em 
\citep{Baron2004,Dessart2006}, SN2005cs 
\citep{Baron2007,Dessart2008} or 
SN2006bp \citep{Dessart2008}, have been 
studied using detailed radiative transfer 
models, which provide a direct way 
to infer information about the chemical 
composition, the density profile and other 
parameters from the full spectral time series.
This applies in particular to the use of SNe~II as 
distance indicators, despite the fact that an 
absolute distance estimate is a natural byproduct 
of a quantitative spectroscopic analysis 
\citep[see][]{Baron1995,Baron1996,Baron2004,Baron2007,Lentz2001,Mitchell2002}.
SNe~II have a long history as cosmological probes 
\citep{Kirshner1974,Schmidt1994} and
have regained popularity in recent years due to the 
increased availability of data at high redshifts 
\citep[e.g.,][]{Poznanski2009,Gall2017} and, in the 
era of high precision cosmology, due to the rising 
need for independent tests of our cosmological 
models. Recent efforts have focused mainly on 
methods that rely on various observed correlations 
between photometric and spectroscopic parameters 
such as the standard candle method (SCM) by 
\citet{Hamuy2002}, the photometric color
method by \citet{DeJaeger2015} or the photospheric 
magnitude method by \citet{Rodriguez2014}. 
Both \citet{Gall2017} and \citet{DeJaeger2017} demonstrate
that with the SCM or the expanding 
photosphere method (EPM) distance measurements of 
SNe~II up to redshifts of $\approx 0.34$ are 
feasible, highlighting the progress that has been 
made possible through the availability of new data. 
In contrast, the determination of distances from 
radiative transfer modeling has stagnated in 
recent years. Both the tailored EPM technique 
\citep{Dessart2006,Dessart2008} and the spectral 
fitting expanding atmosphere method (SEAM) 
\citep{Baron1995,Baron1996,Baron2004,Baron2007} 
have never been applied outside the local universe. 
Nevertheless, their independence of the cosmic 
distance ladder as well as their foundation in 
well-understood physics make them a promising 
independent tool for cosmology.

Motivated by the wealth of available spectral data 
and the unique diagnostic abilities of radiative 
transfer modeling, we have developed a new 
numerical tool for performing spectroscopic 
analysis of SNe~II. Since our main goal is to provide 
a tool for parameter inference, we neglect time-dependent 
effects in favor of computational expediency. Currently, 
the high computational costs prevent the application of 
numerical methods that self-consistently simulate the 
time evolution of the radiation field and the plasma state 
based on initial conditions \citep{Dessart2011,Dessart2013} 
to this purpose. Our approach is an 
extension of the Monte Carlo radiative transfer 
code \textsc{tardis} \citep{Kerzendorf2014}, which 
was originally developed for spectral synthesis in 
type Ia supernovae (SNe~Ia). We have extensively 
modified and improved the physical treatment of 
radiative transfer implemented in \textsc{tardis} 
to be applicable to the modeling of SNe~II 
atmospheres. This improved version of the code is 
then used to calculate a new, independent set of 
dilution factors for the EPM technique. In the EPM 
the dilution factors as introduced by 
\citet{Hershkowitz1986a,Hershkowitz1986b} and 
\citet{Hershkowitz1987} correct for the deviation 
of the supernova emission from that of a blackbody 
of the same color temperature.
They provide the possibility to compare our model 
calculations to previously published numerical 
results by \citet[E96 from now on]{Eastman1996} and 
\citet[D05 from now on]{Dessart2005a} in a simple 
parametrized fashion. Currently, the systematic 
discrepancies between the two sets of dilution 
factors constitute one of the most significant 
sources of uncertainty in the EPM method, 
accounting for differences of roughly 20\% in the 
inferred distance
\citep[e.g.,][]{Takats2006,Jones2009,Gall2016,Gall2017}. 
This significant uncertainty highlights the need 
for additional calculations based on independent 
numerical methods to understand and resolve the 
current tension. 

The structure of the paper is as follows.
We begin with a detailed description of the 
physical extensions and their numerical 
implementation into \textsc{tardis} in 
\cref{sec:Method}. In \cref{sec:EPM}, we provide a 
brief review of the EPM and discuss the basic 
physics of the dilution factors. As a first 
application of the extended version of 
\textsc{tardis}, we present radiative transfer 
models for two epochs of the prototypical 
SN~II SN1999em in \cref{sec:example_spectra}. The 
next sections are dedicated to the presentation 
and discussion of our main application, the 
calculation of a new set of EPM dilution factors. 
The setup of the necessary grid of supernova models 
is described in \cref{sec:ModelGrid}, followed by 
an analysis of the calculated dilution factors in 
\cref{sec:Results}. Here, we focus particularly on 
the differential influence of the model parameters 
such as photospheric density or metallicity. To put 
our results into context and to understand the 
differences between the published set of dilution 
factors, a comparison to previous studies is given 
in \cref{sec:prev_studies}. We investigate the 
differences in the adopted numerical approaches and 
examine the different choices for the 
atmospheric properties. Finally, we summarize our 
results and give an outlook 
in \cref{sec:Conclusions}.

\section{Method} \label{sec:Method}
We present an extended version of the 
one-dimensional Monte Carlo (MC) radiative transfer 
code {\sc tardis} \citep{Kerzendorf2014} that has 
been significantly extended for the application to 
SNe II. {\sc tardis} is based on the indivisible 
energy packet MC methods of
\citet{Lucy1999b,Lucy1999a,Lucy2002,Lucy2003} and 
has been developed for rapid spectral modeling of 
\snias{}. It has been used to study
various aspects of \snia{} explosion physics. 
Applications include abundance
tomographies \citep{Barna2017}, a study of spectral 
signatures of helium in double-detonation models 
\citep{Boyle2017}, as well as analyses of \sniasx{}
spectra \citep{Magee2016,Magee2017}.  In these 
studies only the effects of Thomson scattering 
and bound-bound line interactions are simulated in 
detail. This is a reasonable approximation for 
SNe~Ia but not for SNe~IIP, which have 
a higher ratio of continuum to line opacity due 
to the hydrogen-rich composition. 
To adapt {\sc tardis} to these conditions, we 
extend our treatment of radiation--matter 
interactions to include bound-free, free-free as 
well as collisional processes using the macro atom 
scheme of \citet{Lucy2002,Lucy2003} as outlined 
in \cref{sec:MC}. Further necessary improvements
to the code can be motivated based on the
peculiarities of radiative transfer in SNe~II. 
SNe~II atmospheres are characterized by 
comparatively low densities at the photosphere and 
a scattering dominated opacity.  Due to the low
densities, collisions are ineffective at 
coupling the level populations and 
ionization and excitation are mainly controlled by 
the radiation field. The radiation field is dilute
compared to its equilibrium value as a result of 
the dominance of electron-scattering opacity and 
thus significant departures from local
thermodynamic equilibrium (LTE) arise even far 
below the photosphere \citep[see e.g.,][]{Dessart2011}. 
To address this issue, we have extended the code as
outlined in \cref{sec:NLTE}.  Another consequence 
of the scattering dominated environment is that 
relatively high optical depths on the order of
$\tau \propto \mathcal{O}(10)$ are needed to 
guarantee a full thermalization of the radiation
\citep[see e.g.,][]{Eastman1996}. At such high 
optical depths the atmospheric structure is 
strongly affected by relativistic transfer effects 
as demonstrated by
\citet{HauschildtP.H.BestM.Wehrse1991RelativisticPhotospheres}. 
The inclusion of these effect in \textsc{tardis} is 
described in \cref{sec:RelTrans}.

\subsection{Monte Carlo simulations} \label{sec:MC}
To find a consistent solution for the plasma state 
and the radiation field, \textsc{tardis} performs a 
series of Monte Carlo radiative transfer
simulations.  At every radiative transfer step, a 
large ensemble of indivisible energy packets 
\citep[see][]{Abbott1985,Lucy1999b,Lucy2002,Lucy2003}
representing monochromatic photon bundles is 
initialized at the inner boundary.
Initial packet properties are assigned under the 
assumptions of the LTE diffusion limit. 
Thus, packet frequencies are sampled from a blackbody 
distribution at the inner boundary temperature 
$T_\mathrm{i}$ and propagation directions are 
selected according to zero limb-darkening in the 
comoving frame. Uniform packet energies are chosen such 
that the injected packets carry a total comoving frame luminosity
$L_\mathrm{i} = 4 \pi R_\mathrm{i}^2 \sigma T_\mathrm{i}^4$, 
where $\sigma$ is the Stefan-Boltzmann constant and $R_\mathrm{i}$ 
is the radius of the inner boundary. With initial properties 
assigned, the propagation of the packets is simulated under 
the assumption of a steady-state,
that is to say, neglecting time dependence, as outlined in 
the following section.

\subsubsection{Packet propagation}
After initialization, each packet is followed until 
it leaves the computational domain through the 
inner or outer boundary.  Between the boundaries 
the supernova atmosphere has been discretized into 
equidistant, spherical shells.
Within each shell the plasma properties such as the 
opacity or the electron temperature are assumed to 
be constant. During the propagation the effects of 
Thomson scattering, hydrogen bound-free,
free-free, bound-bound as well as collisional 
processes are taken into account.
As described in \citet{Kerzendorf2014}, line 
opacity is treated in the Sobolev
approximation \citep[see][]{Sobolev1957}. For 
micro-turbulent velocities on the order of 
$\SI{100}{\km \per \second}$, this is as accurate 
as the comoving frame method in describing the 
formation of the Balmer lines in
SNe~IIP \citep[see][]{Duschinger1995f}. Following 
\citet{Lucy2003}, the free-free opacity
\begin{equation} \label{eq:ff_opacity}
  \chi_\mathrm{ff} (\nu) =  \alpha_\mathrm{ff} \, 
  \nu^{-3} T_\mathrm{e} ^{-1/2} 
  n_\mathrm{e} \left( 1 - e^{-h\nu/k_\mathrm{B} T_\mathrm{e}} \right) 
  \sum_{j,k} N_{j,k} (j-1)^2  
\end{equation}
is evaluated with free-free gaunt factors set to 
unity.  Here, $N_{j,k}$ denotes the number density 
of ionization stage $j$ of element $k$,
$T_\mathrm{e}$ is the electron temperature, 
$n_\mathrm{e}$ the electron density and $\nu$ the 
frequency. The prefactor $\alpha_\mathrm{ff}$ has the 
value \SI{3.69e8}{\cm^5 \kelvin^{1/2} \per s^3}. 
Since hydrogen is the dominant source of 
bound-free opacity in \sniis{}, we restrict the 
inclusion of these processes currently only to this
element. However, since an extension to more 
species is conceptually straightforward, we present 
the governing equations in their
general form. Thus, the opacity resulting from 
photoionizations of electrons in level
$i$ of ion $j,k$ is given by
\begin{equation} \label{eq:BfOpacity}
  \chi_{i,j,k}^\mathrm{bf} (\nu) = 
  \alpha_{i,j,k \rightarrow j+1, k} (\nu)
    \left( n_{i,j,k} - n_{i,j,k}^*  
    e^{-h\nu/k_\mathrm{B} T_\mathrm{e}} \right),
\end{equation}
where $n_{i,j,k}$ and $n_{i,j,k}^*$ denote the 
actual and the respective LTE level number densities 
\citep[see equation 5.25 of ][]{hubeny2014}.  The 
cross-section for photoionzation $\alpha_{i,j,k
\rightarrow j+1, k}(\nu)$ is obtained from 
tabulated values through linear interpolation. 

To account for the inclusion of hydrogen 
bound-free, free-free as well as collisional 
processes small modifications to the packet 
propagation procedure of \citet[\textsection
2.6]{Kerzendorf2014} have been necessary.  In 
particular for continuum interactions an additional 
MC experiment is needed to determine the physical
absorption mechanism. The probabilities for bound-
free absorption, free-free absorption and Thomson 
scattering are given by 
$\chi^\mathrm{bf}/(\chi^\mathrm{bf} + \chi^\mathrm{ff} + 
\chi^{\mathrm{Th}})$, $\chi^\mathrm{ff}/(\chi^\mathrm{bf} + 
\chi^\mathrm{ff} + \chi^{\mathrm{Th}})$ and 
$\chi^\mathrm{Th}/(\chi^\mathrm{bf} + 
\chi^\mathrm{ff} + \chi^{\mathrm{Th}})$ 
respectively. If a bound-free process is selected, 
a specific continuum for absorption has to
be assigned according to the probabilities 
$\chi^\mathrm{bf}_{i,j,k}/
\chi^\mathrm{bf}$ for photoionization from specific 
levels $i$ of ion $j, k$. Regardless of the type of 
interaction, we use the macro atom scheme of 
\citet{Lucy2002,Lucy2003} to select an 
emission channel as outlined in 
\cref{sec:macro_atom}.  For bound-free and 
free-free emission, the packet has to be assigned 
an appropriate frequency before the
propagation can be resumed. We 
employ the approximate sampling rule of 
\citet[Eq. 41]{Lucy2003} for free-free processes 
and linear interpolation on precomputed values of 
the emissivity for bound-free interactions.

\subsubsection{Macro atom} \label{sec:macro_atom}
We use the macro atom scheme of 
\citet{Lucy2002,Lucy2003} for a general treatment 
of complicated radiation-matter interactions, such 
as recombination cascades, fluorescent line 
emission or cooling emission.  In this scheme, 
packet splitting for processes with multiple 
emission channels is avoided by assigning
the total energy of the packet randomly to one 
possible interaction channel according to a set of 
rules derived from the assumption of statistical
equilibrium. In \citet{Kerzendorf2014}, only the 
redistribution of excitation energy created
by bound-bound absorption events was simulated 
using the macro atom machinery. We introduce 
indivisible packets of thermal kinetic energy ($k$-
packets) and ionization energy ($i$-packets) in 
addition to the packets of excitation energy
(macro atoms) included in \citet{Kerzendorf2014}
to treat continuum interactions.  $k$-packets can 
be created by bound-free and free-free absorption 
events as well as collisional deactivations of 
$i$-packets or macro atoms. Since both thermal and 
ionization energy are created in photoabsorption 
events, the $r$-packet is transformed into a
$k$-packet with probability 
$p^k=\nu_{i,j,k}/\nu^\prime$ and into an $i$-packet
otherwise. Here, $\nu_{i,j,k}$ is the threshold for 
ionization and $\nu^\prime$ is the frequency of the 
$r$-packet in the comoving frame. Based on the assumption of radiative 
balance in the fluid rest frame, all $i$-packets, 
macro atoms and $k$-packets have to be converted
in-situ back to $r$-packets. 
For $k$-packets this is done by sampling the rates 
at which different physical processes cool the 
electron gas.  All treated cooling rates are listed 
in \cref{sec:thermal_balance}.  For macro atoms and 
$i$-packets, the situation is more complicated due 
to the possibility of internal transitions. In both 
cases we sample the internal energy flow rates 
until a radiative deexcitaton process is selected 
or a collisional deactivation to a $k$-packet 
occurs \citep[see][]{Lucy2002,Lucy2003}. 
The needed energy flow rates are calculated
with rate coefficients evaluated as described in 
\cref{sec:NLTE}.

\subsubsection{Reconstruction of radiation field quantities} 
\label{sec:Estimators}
For our detailed treatment of ionization and 
thermal structure (see \cref{sec:NLTE}), estimates 
for the radiative bound-free rates and radiative
heating rates are needed. We use volume-based 
estimators \citep{Lucy1999a,Lucy2003} to
reconstruct the relevant quantities from the 
trajectories of the packet ensemble.  In this 
approach, the time-averaged contributions of all 
trajectory segments, on which the process can in 
principle occur, are taken into account.
Thus, to obtain an estimate for the photoionization 
rate coefficient $\gamma_{i,j,k}$ for level 
$i,j,k$, we sum over all path segments $\dif{s}$ 
for which the comoving frame (CMF) frequency 
$\nu^\prime$ of the packet is larger
than the threshold for photoionization 
$\nu_{i,j,k}$
\begin{equation} \label{eq:EstimatorPhotoIon}
  \gamma_{i,j,k} = \frac{1}{\Delta t V} \sum_{\nu^\prime \geq \nu_{i,j,k}}
    \varepsilon^\prime_{\nu} \frac{\alpha_{i,j,k \rightarrow j+1,k}(\nu^\prime)}{
      h \nu^\prime} \dif{s}.
\end{equation}
Here, $V$ is the volume of the respective grid
cell and $\varepsilon^\prime_{\nu}$ is
the CMF packet energy.  The time interval $\Delta 
t$ is a numerical normalization factor that is 
determined by the energy injection rate at the
lower computational boundary.
Similarly, the estimator for the stimulated 
recombination rate coefficient is given by
\begin{equation} \label{eq:EstimatorStimRecomb}
  \alpha_{i,j,k}^\mathrm{stim} = 
  \Phi_{i,j,k}(T_{\mathrm{e}})
  \frac{1}{\Delta t V} 
    \sum_{\nu^\prime \geq \nu_{i,j,k}} \epsilon^\prime_{\nu} 
      \frac{\alpha_{i,j,k \rightarrow j+1,k}(\nu^\prime)}{h \nu^\prime} 
        e^{-h \nu^\prime /k_\mathrm{B} T_\mathrm{e}} \dif{s}.
\end{equation}
Here, the Saha factor
\begin{equation} \label{eq:SahaFactor}
  \Phi_{i,j,k}(T) = \frac{n_{i,j,k}^*}{n_{0,j+1,k}^* n_\mathrm{e}}
\end{equation}
enters, which connects the LTE level populations 
$n_{i,j,k}^*$ to the ground state population
$n_{0,j+1,k}^*$ of the next higher ionization 
stage. The heating rate coefficient for 
photoionization is
\begin{equation}
  h^\mathrm{bf}_{i,j,k} = \frac{1}{\Delta t V} \sum_{\nu^\prime \geq \nu_{i,j,k}}
    \varepsilon^\prime_{\nu} \alpha_{i,j,k \rightarrow j+1,k}(\nu^\prime)
      \left(1 -  \frac{\nu_{i,j,k}}{\nu^\prime} \right) \dif{s}.
\end{equation}
Finally, the heating rate $H^\mathrm{ff}$ due to 
inverse-bremsstrahlung is calculated using
\begin{equation}
  H^\mathrm{ff} = \frac{1}{\Delta t V} \sum \chi_\mathrm{ff}(\nu) 
  \epsilon^\prime_{\nu} \dif{s},
\end{equation}
with the free-free opacity $\chi_\mathrm{ff}(\nu)$ 
treated according to \Cref{eq:ff_opacity}. 
Before concluding our presentation of the 
reconstruction of radiation field quantities, 
we stress again that currently the estimators for 
the bound-free processes $\gamma_{i,j,k}$,
$\alpha_{i,j,k}^\mathrm{stim}$
and $h^\mathrm{bf}_{i,j,k}$ are only used for 
hydrogen.

\subsubsection{Relativistic transfer} \label{sec:RelTrans}
For photospheric-phase \sniis{} 
the emergent continuum radiation is created in 
regions well below the photosphere. In these 
optically thick regions, the radiation field is
essentially isotropic in the fluid rest frame and 
relativistic frame transformations can 
significantly modify the energy transport in the 
ejecta by introducing small anisotropies in the lab 
frame intensity \citep[see
e.g.,][]{HauschildtP.H.BestM.Wehrse1991RelativisticPhotospheres,Baron1996a}. 

To include relativistic effects in the Monte Carlo 
simulations, \textsc{tardis} uses a mixed-frame 
approach. Radiation--matter interactions are 
handled in the comoving frame whereas the packet 
propagation is carried out in the lab frame.
Whenever necessary we transform the relevant packet 
properties between the frames. Compared to 
\citet{Kerzendorf2014}, we 
have refined the treatment of relativity by 
including frame transformations of opacities as 
well as angle aberration. To transform packet 
energies and frequencies between observer and
comoving frame, we use the full Doppler factor 
instead of an first order approximation. 
Expressions for the relevant transformations laws 
can be found in \citet{Mihalas1984} or, 
specifically for spherical geometries, in
\citet{Castor1972}. To be consistent with the 
adopted frame transformations,
the distance to the next possible line interaction 
is now calculated based on the full Doppler-shift 
formula. As a result, the common-direction 
frequency surfaces, that is, the surfaces that emit 
line radiation at the same frequency in the 
observer frame, are no longer planes perpendicular 
to the line of sight but have a more complicated 
geometry as described by the relativistic Sobolev 
theory of \citet{Jeffery1995}.

\subsection{Plasma state} \label{sec:NLTE}
The original implementation of \textsc{tardis} only 
features approximate excitation and ionization 
treatments and a very simplified calculation of the
thermal structure. We have considerably refined the 
determination of the plasma state to adapt the code 
to \sniis{}. 
In particular, we have implemented a full NLTE 
treatment of excitation and ionization for hydrogen 
and we employ a thermal balance calculation 
to infer the temperature structure of the envelope. 

The calculation of the plasma state involves a 
simultaneous determination of the excitation and 
ionization state of the material as well as the 
thermal structure. To reduce the complexity of this 
nonlinear problem, we decouple the
solution of the excitation and ionization balance 
as follows: given an initial guess for the kinetic 
temperature $T_{\mathrm{e}}$ and the electron 
density $n_{\mathrm{e}}$, we calculate level 
population fractions as outlined in
\cref{sec:exc}. Based on the obtained excitation 
state, we solve for the ionization balance as 
described in \cref{sec:Ion}. Finally, we compute 
heating and cooling rates (see 
\cref{sec:thermal_balance}), which are needed for 
the determination of the thermal structure.  An 
outer iteration loop establishes consistency 
between excitation and ionization and adjusts the
temperature such that thermal balance is enforced 
(see \cref{sec:plasma_outer}).

\subsubsection{Excitation} \label{sec:exc}
{\sc tardis} offers excitation treatments with 
different levels of sophistication.  Level 
population fractions 
$f_{i,j,k} = n_{i,j,k} / N_{j,k}$
can be calculated from the Boltzmann excitation 
equation, a nebular modification thereof 
\citep[see][]{Abbott1985} or from the steady-state
equations of statistical equilibrium.

For the NLTE excitation calculation, electron 
number densities have to be specified. In this 
case, the statistical equilibrium
equations for the total system decouple and can be 
solved for each atomic species individually. 
In the NLTE treatment of \citet{Kerzendorf2014}, 
only bound-bound interactions
and collisional excitation and deexcitation rates 
were included. We extend the scheme by including 
radiative and collisional bound-free processes to 
obtain a more complete description of hydrogen 
excitation.  The necessary photoionization and recombination 
rate coefficients $\gamma_i$ and $\alpha_i$ are 
reconstructed from the MC simulation by volume-
based estimators (see \cref{sec:Estimators}). The
collisional ionization and recombination rate 
coefficients $q_{i \kappa}$ and
$q_{\kappa i}$ are evaluated according to the 
approximate formula by \cite{Seaton1962}. 
With these processes included, the rate equation 
for level $i$ of ion $j,k$ is given by
\begin{equation} \label{eq:NLTE_exc}
- \left(\gamma_{i} + q_{i \kappa} n_\mathrm{e} + 
\sum_{m \neq i} r_{im} \right) f_{i} + 
\sum_{m \neq i} r_{mi} f_m = -(\alpha_{i} + 
q_{\kappa i} n_\mathrm{e}) \frac{ N_{j+1,k} n_\mathrm{e}}{N_{j,k}}.
\end{equation}
Here, $r_{mi}$ and $r_{im}$ denote the total rate 
coefficients at which radiative and collisional 
transitions between level $i$ and $m$ populate and
depopulate level $i$. In the Sobolev approximation, 
the rate coefficient for
deexcitation from an upper level $u$ to a lower 
level $l$ is given by
\begin{equation}
 r_{ul} = \beta_{lu} A_{ul} + \beta_{lu} B_{ul} 
 J^\mathrm{b}_{lu} + c_{ul} n_\mathrm{e}
\end{equation}
and the excitation rate coefficient is
\begin{equation}
 r_{lu} = \beta_{lu} B_{lu} J^\mathrm{b}_{lu} + c_{lu} n_\mathrm{e}.
\end{equation}
Here, $J^\mathrm{b}_{lu}$ is the mean intensity at 
the blue wing of the line,
$\beta_{lu}$ is the Sobolev escape probability 
\citep[see e.g., \textsection
4.3.1 of][]{Lucy2002} and $A_{lu}, B_{lu}$ and 
$B_{ul}$ are the Einstein
coefficients. Electron impact excitation rate 
coefficients $c_{lu}$ are taken
from the approximate formula of 
\citet{VanRegemorter1962} with deexcitation
rates evaluated according to detailed balance. For 
hydrogen levels with principal quantum numbers up 
to $n=7$ we use collision strengths from the
detailed ab initio calculations of 
\citet{Przybilla2004}.

Despite fixing the electron number densities, the 
system of rates equations \labelcref{eq:NLTE_exc} 
remains nonlinear due to the dependence of the 
Sobolev escape probabilities on the level 
populations.  We use a standard root finding
algorithm to solve for the level population 
fractions 
$f_{i,j,k} = n_{i,j,k} / N_{j,k}$ and the ion 
population ratio $N_{j+1,k}/N_{j,k}$.
\footnote{Specifically, we use a modified version of the 
Powell hybrid method as implemented in \textsc{SciPy} 
\citep{Scipy}.} The convergence of the outer iteration loop 
that establishes a consistent plasma state is
accelerated considerably by including the
ion population ratio in the solution of the 
excitation state.

\subsubsection{Ionization} \label{sec:Ion}
In our detailed treatment of ionization we use the 
derived level population
fractions $n_{i,j,k} / N_{j,k}$ and an initial 
guess for the electron density
$n_{\mathrm{e}}$ to calculate the total ionization 
rate coefficient
\begin{equation}
  \Gamma_{j,k} = \sum_{i} n_{i,j,k} (q_{i,j,k \rightarrow j+1,k} 
  n_\mathrm{e} + \gamma_{i,j,k}) / N_{j,k}
\end{equation}
and the total recombination rate coefficient
\begin{equation}
\alpha_{j+1,k} = \sum_{i} \left( \alpha^\mathrm{sp}_{i,j,k} + 
\alpha^\mathrm{stim}_{i,j,k} + q_{i,j,k \leftarrow j+1,k} 
n_\mathrm{e} \right)
\end{equation}
for relevant pairs of ions $(j,k)$, $(j+1,k)$.  We 
only do this for hydrogen in this work.
For all other ions, the Saha factor
$\Phi_{j,k} = (N_{j+1,k} n_\mathrm{e}) / N_{j,k}$ 
and the electron density $n_\mathrm{e}$ serve as 
approximations of the total ionization and
recombination rate coefficients. The Saha factor 
$\Phi_{j,k}$ is evaluated according to the Saha 
equation at the local radiation temperature
$T_\mathrm{R}$ or the nebular ionization formula of 
\citet{Mazzali1993}
\citep[see Eqs. 2 and 3 of][]{Kerzendorf2014}.
Based on these ionization and recombination rate 
coefficients, we iteratively solve for 
the ion and electron number densities, assuming 
ionization equilibrium. 

\subsubsection{Thermal balance} \label{sec:thermal_balance}
To complete the description of the plasma state, we 
need an estimate for the electron temperature 
$T_\mathrm{e}$ in the ejecta. In
\citet{Kerzendorf2014}, $T_\mathrm{e}$ was set to 
$0.9 T_\mathrm{R}$ following \citet{Mazzali1993}. 
We replace this simplified treatment of the
thermal structure by a thermal-balance 
calculation based on the heating and cooling rates 
of the gas.
 
The thermal energy content,
and therefore the temperature, is determined by the 
energy exchange between the kinetic energy of the ejecta, the
radiative energy pool and the pool of atomic 
internal energy. This transfer is
mediated by adiabatic cooling, collisional transitions as well as 
bound-free and free-free interactions.
Assuming a steady-state, the rates for heating and 
cooling of the ejecta by these processes must 
cancel. Thus the electron temperature 
$T_\mathrm{e}$ is fixed by the requirement of 
thermal balance
\begin{equation} \label{eq:thermal_balance}
H^\mathrm{bf} + H^\mathrm{ff} + H^\mathrm{deexc} + H^\mathrm{recomb} 
= C^\mathrm{fb} + C^\mathrm{exc} + C^\mathrm{ion} + C^\mathrm{ad} + 
C^\mathrm{ff}.
\end{equation}
Here, $H^\mathrm{bf}$ and $C^\mathrm{fb}$ denote 
the rates of heating and cooling by bound-free 
interactions, $H^\mathrm{ff}$ and $C^\mathrm{ff}$ 
the respective rates for free-free processes. The 
contributions from collisional
excitation, deexcitation, ionization and 
recombination are $C^\mathrm{exc}$,
$H^\mathrm{deexc}$, $C^\mathrm{ion}$ and 
$H^\mathrm{recomb}$. The final term
$C^\mathrm{ad}$ describes adiabatic cooling of the 
envelope due to expansion work.

Specifically, collisional excitations from lower 
levels $l,j,k$ to level $i,j,k$ remove energy from 
the thermal pool with a rate
\begin{equation} \label{eq:CollExecCooling}
  C_{i,j,k}^\mathrm{exc} = \sum_l C_{l,j,k \rightarrow i,j,k}
    (\epsilon_{i,j,k} - \epsilon_{l,j,k}),
\end{equation}
where $\epsilon_{i,j,k}$ and $\epsilon_{l,j,k}$ are 
the respective level energies. Correspondingly, 
collisional ionizations from bound levels of ion
$j,k$ contribute
\begin{equation}
  C_{j+1,k}^\mathrm{ion} = \sum_l C_{l,j,k \rightarrow j+1,k}
    (\epsilon_{0,j+1,k} - \epsilon_{l,j,k})
\end{equation}
to the total cooling rate.

In turn, atomic internal energy is transfered to 
the thermal pool by the inverse processes of 
collisional recombination and deexcitation at rates
\begin{equation}
 H_{j,k}^\mathrm{recomb} = \sum_l C_{l,j,k \leftarrow j+1,k}
    (\epsilon_{0,j+1,k} - \epsilon_{l,j,k})
\end{equation}
and
\begin{equation} \label{eq:CollDeexecHeating}
  H_{i,j,k}^\mathrm{deexc} = \sum_u C_{i,j,k \leftarrow u,j,k}
    (\epsilon_{u,j,k} - \epsilon_{i,j,k}).
\end{equation} 
Thermal electrons moving in the field of an ion 
$j,k$ emit radiative energy according to
\citep[see][]{Osterbrock1974}
\begin{equation}
  C_{j,k}^\mathrm{ff} = \SI{1.426e-27} (j-1)^2 
  T_\mathrm{e}^{1/2} N_{j,k} n_\mathrm{e},
\end{equation}
which depends on the ionic charge $j-1$, the number 
density of the respective
ion $N_{j,k}$ as well as $T_\mathrm{e}$ and 
$n_\mathrm{e}$. In addition, energy is continuously 
removed from the thermal electron pool by radiative
recombinations. In terms of the modified rate 
coefficient
\begin{equation} \label{eq:RateCoeffSpRecombMod}
  \alpha_{i,j,k}^{\mathrm{E,sp}} = 4 \pi \, \Phi_{i,j,k}(T_{\mathrm{e}})
    \int_{\nu_{i,j,k}}^{\infty} \frac{\alpha_{i,j,k \rightarrow j+1,k}(\nu)}{
      h \nu_{i,j,k}}
        \frac{2 h \nu^3}{c^2} e^{-h \nu/ k_\mathrm{B} T_{\mathrm{e}}} \dif{\nu}
\end{equation}
the cooling rate by 
recombinations to level $i,j,k$ can be written as
\begin{equation}
  C_{i,j,k}^{\mathrm{fb,sp}} = N_{j+1,k} n_\mathrm{e}
    \left(\alpha_{i,j,k}^{\mathrm{E,sp}} - \alpha_{i,j,k}^{\mathrm{sp}} \right)  
      h \nu_{i,j,k}.
\end{equation}
Photoionizations, in turn, heat the medium with a rate 
\begin{equation}
 H_{i,j,k}^\mathrm{bf} = 4 \pi n_{i,j,k}
    \int_{\nu_{i,j,k}}^{\infty} \alpha_{i,j,k \rightarrow j+1,k}(\nu) 
    \left( 1  - \frac{\nu_{i,j,k}}{\nu} \right)
        J_\nu \dif{\nu}.
\end{equation}
Finally, the electron gas continuously loses 
thermal kinetic due to the expansion of the ejecta. 
The rate of energy loss resulting from this 
adiabatic cooling is given by
\begin{equation}
  C^\mathrm{ad} = 3 n_\mathrm{e} k_\mathrm{B} T_\mathrm{e} / t,
\end{equation}
where $t$ denotes the time of explosion. 

\subsubsection{Outer plasma iteration} \label{sec:plasma_outer}
To obtain a consistent solution for the plasma 
state, the input electron densities that are used 
in the calculation of the level population 
fractions have to agree with the results from the 
ionization calculation. This is achieved by 
combining the methods described above with an 
iterative root-finding procedure. Apart from 
establishing consistency between the
excitation and ionization state, the outer 
iteration loop is used to determine
the thermal structure from the thermal balance 
equation \labelcref{eq:thermal_balance}.

\subsection{Approximations} \label{sec:approx}
As established by 
\citet{Utrobin2005,Dessart2008a,Dessart2010,Potashov2017} 
time-dependent effects in the excitation and ionization balance can 
play an important role in shaping the spectral energy distribution.
This applies in particular to epochs following hydrogen recombination. 
At these times the inclusion of time-dependent terms induces an 
overionization compared to the steady state solution, which is crucial 
in reproducing the observed H$\alpha$ line strengths. In contrast, for 
epochs preceding hydrogen recombination the influence of time dependence is 
modest.  Since these epochs are most relevant for the application of 
EPM \citep[see e.g.,][]{Dessart2006,Dessart2008}, we do not consider our 
neglect of these effects a severe limitation to our approach. In fact, 
\citet{Dessart2008a} find negligible differences between the dilution 
factors from their time-dependent calculations and the steady-state 
results from \citet{Dessart2006,Dessart2008}.
Only for color temperatures less than \SI{7000}{\kelvin} the correction 
factors drop systematically below their steady-state counterparts.
As an additional approximation in the solution of 
the statistical equilibrium equations, we assume detailed radiative 
balance in the Lyman continuum. 
This prevents MC noise in the estimator for the ground-
state photoionization rate from hindering 
convergence and avoids associated fluctuations in 
the ionization as well as the heating and
cooling balance.  This approach follows previous 
studies of SNe~IIP, such as
\citet{Takeda1990f,Takeda1991f} and 
\citet{Duschinger1995f}.  Since the Lyman continuum 
is optically thick as long as the outflow 
ionization is not extremely high, detailed balance 
is deemed to be a very good approximation under 
most conditions of interest. 
We have verified this by a series of test 
calculations without this assumption but with 
increased numbers of MC quanta. The spectra 
resulting from the two approaches show good 
agreement for the parameter space that has been 
investigated in this paper.
This is consistent with the results of 
\citet{Duschinger1995f} who have reached the same 
conclusion for pure hydrogen supernova
atmospheres with photospheric temperatures up to 
\SI{15000}{\kelvin}.

\subsection{Iteration cycle}
\begin{figure}
  \begin{center}
  \hspace*{-0.45cm}
  \vspace{-0.45cm}
    \includegraphics[width=0.45\textwidth]{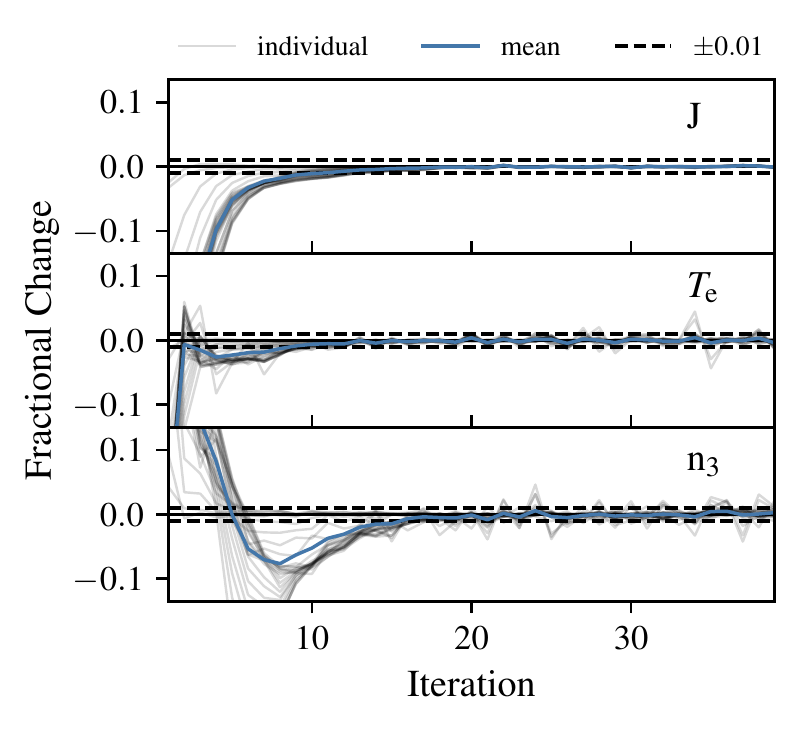}
  \end{center}
  \caption[]{Illustration of the convergence properties of plasma 
  and radiation field quantities. We show the fractional changes between 
  successive iterations for the mean intensity of the radiation field $J$, 
  the electron temperature $T_\mathrm{e}$ and a representative level 
  population, specifically that of the second excited level of hydrogen $n_3$. 
  In all cases, we include both the changes in each individual shell (gray) 
  as well as their average (blue). The results shown here are taken from our 
  SN1999em model for the 14th of November (see \cref{sec:example_spectra} 
  and \cref{fig:sn1999em_1411_comp}).
 }
  \label{fig:Convergence}
\end{figure}

\textsc{Tardis} alternates between the calculation of the plasma 
state and MC radiative steps to achieve a self-consistent state for 
radiation and matter. Generally, less than twenty of these iterations 
are needed to achieve convergence to a point where only statistical 
variations remain (see \cref{fig:Convergence}). The good convergence 
properties result from the strict enforcement of radiative equilibrium, 
the explicit treatment of scattering and the direct dependence of the 
macro atom emissivities on the current estimate of the radiation field 
through the macro atom activation rates. At the moment, the number of 
iterations is set by hand at the beginning of the simulation, since no formal 
convergence criterion is implemented in \textsc{Tardis}. For the calculations 
presented in this paper, we have performed 40 iterations in all cases. 
We have found this to be more than sufficient to guarantee convergence for 
all used setups.

\subsection{Atomic data}
We use the hydrogen atomic data as described by 
\citet{Sim2005}. This data set is based on a 20 level model 
atom with each level corresponding to a
principal quantum number $n$.  Frequency-dependent 
photoionization cross sections are tabulated for 
every energy state. The tabulated values range from 
the threshold ionization frequency up to the
point at which the cross-section is only about 
\SI{0.07}{\percent} of the value
at threshold. This improved hydrogen model atom 
complements the atomic data already included in 
{\sc tardis} \citep[see][]{Kerzendorf2014},
which is compiled from the line lists of 
\citet{Kurucz1995} and the Chianti 7.1
data base \citep{Dere1997,Landi2012}.

\subsection{Spectral synthesis}
To calculate synthetic spectra, the properties of 
escaping packets are recorded and later binned.
However, the quality of the spectra that can be 
obtained from the normal MC quanta is severely 
affected by MC noise. To improve the 
signal-to-noise ratio an additional type of MC 
packet is used in \textsc{tardis}.  Whenever a 
normal packet is launched or performs
an interaction in the final spectral synthesis run, 
these so called ``virtual'' packets ($v$-packets) 
are emitted to estimate the contribution of the 
event to the emergent spectrum. 
In practice this amounts to optical depth 
integrations along a number of
randomly selected trajectories through the ejecta. 
The measured optical depths
$\tau_\mathrm{trj}$ are then used to weight the 
contributions of the virtual
packets to the spectrum according to the escape 
probability along the packet
path, $\exp(-\tau_\mathrm{trj})$. This procedure, 
commonly called ``peeling off'', is well 
established and has found widespread use, in 
particular in the area of dust MC radiative 
transfer \citep[see e.g.,][]{Yusef-Zadeh1984,
Wood1999, Baes2011, Steinacker2013, Lee2017}. 

Compared to the implementation described in 
\citet{Kerzendorf2014}, modifications have been 
necessary to keep the computational effort 
reasonable for the high optical depths of our 
\snii{} atmospheres. Since the number of
interactions scales quadratically with optical 
depth, the number of virtual
packets that have to be tracked for each ``real'' 
packet quickly becomes
prohibitively large as the optical depth is 
increased. At the same time, the
contribution of the additional $v$-packets to the 
spectrum is marginal due to
the strong attenuation towards the surface.  We 
apply biasing to the virtual packet emission to 
tackle this issue. Virtual packets are created only 
with a probability $\exp(-\tau_\mathrm{e})$, where 
$\tau_\mathrm{e}$ is the electron
scattering optical depth.  To account for the lower 
chance of creation, the
weight of the spawned packet is increased by the 
inverse of this probability. 
Notwithstanding this application of biasing, 
virtual packets can still accumulate large amounts 
of optical depth, for example in line interactions. We use 
the Russian roulette technique 
\citep[see e.g.,][]{Carter1975, Dupree2002} to
probabilistically remove these low-weight packets.

\subsection{Supernova model} \label{sec:SNModel}
{\sc tardis} allows for the use of complex 
supernova models based on hydrodynamical explosion 
simulations and with stratified abundances
\citep[see][Appendix A]{Kerzendorf2014}.  
Nevertheless, to facilitate the
exploration of the parameter space, we restrict our 
analysis to simple, highly parameterized models.  
As in \citetalias{Dessart2005a}, we assume power-
law density profiles 
\begin{equation}
   	\rho(r) = \rho_0 (r/r_0)^{-n}
    \label{eq:sn_density_profile}
\end{equation}
with density indexes 
$n=-\dif{\ln{\rho}}/\dif{\ln{r}}$ in the
range $n = 6 - 14$. Both hydrodynamic simulations
\citep{Chevalier1976,Chevalier1982,Blinnikov2000} and spectral
modeling \citep[see
e.g.,][]{Eastman1989,Schmutz1990,Baron2007,Dessart2006,Dessart2008}
have demonstrated that the outer density 
distribution is well described by such
an ansatz with values close to $n \sim 10$.  The 
composition of the ejecta is taken to be 
homogeneous. Heavy elements up to nickel are 
included in the simulations.  Following 
\citetalias{Dessart2005a} we use CNO-cycle 
equilibrium values from \citet{Prantzos1986} for 
the abundances of H, He, C, N, O.  The
remaining elements are assumed to have solar 
chemical composition with values
taken from \citet{Asplund2009}.

\section{Expanding photosphere method} \label{sec:EPM}
\subsection{Presentation of the method}
The expanding photosphere method (EPM) of 
\citet{Kirshner1974} is based on a
simplified model of the supernova 
as a sharply-defined,
spherically-symmetric, expanding photosphere. The 
radiation emerging from this
photosphere is assumed to be that of a blackbody, 
diluted by an amount given by
the dilution factor $\xi_\nu$.  This correction 
factor $\xi_\nu$ has originally been introduced by 
\citet{Hershkowitz1986a,Hershkowitz1986b} and 
\citet{Hershkowitz1987} to correct for the dilution 
of continuum flux that occurs in a
scattering-dominated environment. 
In practice, the dilution factors account for all 
deviations of the spectrum from blackbody emission, 
such as lines or limb-darkening, in a parametrized
fashion \citepalias[see e.g.,][]{Eastman1996,Dessart2005a}. 
For reasons of simplicity, in the application of 
EPM it is assumed that the dilution factor only 
depends on the color temperature. The precise form 
of this dependence may be reconstructed from 
supernovae whose distance is known
from independent means \citep[see][]{Schmidt1992}. 
However, to determine absolute distances it is 
necessary to infer the dilution factors from
theoretical models as in \citetalias{Eastman1996} and 
\citetalias{Dessart2005a}, and outlined 
at the end of this section. 

Based on the assumptions given above, the specific 
luminosity of the supernova is given by
\begin{equation} \label{eq:L_dilute_bb}
  L_\nu = 4 \pi \xi_\nu^2 R_\mathrm{ph}^2 \pi B_\nu(T),
\end{equation} 
where $R_\mathrm{ph}$ is the photospheric radius 
and $T$ is the temperature
of the blackbody $B_\nu(T)$. By equating this to 
the observed de-reddened luminosity 
$L_\nu^\mathrm{obs} = 4 \pi D^2 f_\nu^\mathrm{dered}$ the angular
size of the expanding photosphere
\begin{equation} \label{eq:AngularSize}
  \theta = \frac{2 R_\mathrm{ph}}{D} = 2 \,\,
  \sqrt[]{\frac{f_\nu^\mathrm{dered}}{\xi_\nu^2 \pi B_\nu(T)}}
\end{equation}
can be inferred from the measured de-reddened flux 
$f_\nu^\mathrm{dered}$. The temperature $T$ has to 
be determined from photometry as will be outlined
shortly. Finally, to obtain the distance to the 
supernova
\begin{equation} \label{eq:RphotTheta2Distance}
  D = \frac{2 R_\mathrm{ph}}{\theta},
\end{equation}
the photospheric radius must be eliminated from the 
equations. For homologous expansion this can be 
achieved via the relation
\begin{equation} \label{eq:Rphot}
  R_\mathrm{ph} = v_\mathrm{ph} (t-t_0),
\end{equation}
where $t_0$ is the time of explosion and 
$v_\mathrm{ph}$ is the photospheric
velocity. 
The expansion velocity $v_\mathrm{ph}$ can be 
inferred from blueshift velocities of lines, from 
cross-correlation of the observations with model
spectra \citep[see][]{Hamuy2001} or from tailored 
radiative transfer calculations 
\citep[see e.g.,][]{Dessart2006,Dessart2008}.  
Finally, by measuring the ratio of photospheric 
angular diameter and velocity
\begin{equation}
  \frac{\theta}{v_\mathrm{ph}} = \frac{t-t_0}{D}
\end{equation}
for multiple epochs $t$, the distance is obtained 
from the slope of the data points. The time of 
explosion follows from the intercept with the $t$-
axis.

To apply this formalism 
to observations, we have to recast the relevant 
equations in terms of photometric magnitudes. For a 
bandpass $\bar{\nu}$ with a transmission function
$\phi_{\bar{\nu},\nu}$ 
the apparent magnitude $m_{\bar{\nu}}$ of the 
object can be calculated from the
observed flux $f_\nu^\mathrm{obs}$ according to
\begin{equation}
  m_{\bar{\nu}} = -2.5 \log \left( \int_0^\infty \dif{\nu} 
  \phi_{\bar{\nu}, \nu} f_\nu^\mathrm{obs} \right) + C_{\bar{\nu}},
\end{equation}
where $C_{\bar{\nu}}$ is the zero-point. Using this 
definition, we can rewrite
\cref{eq:L_dilute_bb} for the dilute-blackbody 
emission as follows:
\begin{equation}
  m_{\bar{\nu}} = - 5 \log(\xi) - 5 \log(\theta)  + 
  A_{\bar{\nu}} + b_{\bar{\nu}}
\end{equation}
Here, we have introduced the broadband dust 
extinction $A_{\bar{\nu}}$ and the
blackbody magnitude
\begin{equation}
  b_{\bar{\nu}} = -2.5 \log \left( \int_0^\infty \dif{\nu} 
  \phi_{\bar{\nu}, \nu} \pi B_\nu(T_S) \right) + C_{\bar{\nu}},
\end{equation}
where $T_S$ is the color temperature. By minimizing 
the difference between observed and model 
magnitudes
\begin{equation} \label{eq:ObsEpm}
 \mathcal{E} = \sum_{\bar{\nu} \epsilon S} \big( m_{\bar{\nu}} - 
 A_{\bar{\nu}} + 5 \log (\theta \xi_S) - 
  	b_{\bar{\nu}}(T_S) \big)^2
\end{equation}
for a bandpass combination $S$, the angular 
diameter $\theta$ and the color
temperature $T_S$ can be inferred from photometric 
observations. 
 
To determine dilution factors from a synthetic 
spectrum, \cref{eq:ObsEpm} is
rewritten in terms of absolute magnitudes 
$M_{\bar{\nu}}$:
\begin{equation} \label{eq:SynEpm}
 \mathcal{E} = \sum_{\bar{\nu} \epsilon S} \left( M_{\bar{\nu}} + 
 5 \log \xi_S + 5 \log \frac{R_\mathrm{ph}}{\SI{10}{pc}} - 
  	b_{\bar{\nu}}(T_S) \right)^2
\end{equation}
In this case, the photospheric radius 
$R_\mathrm{ph}$ is known and
application of the minimization procedure to the 
synthetic magnitudes $M_{\bar{\nu}}$ yields the 
color temperature $T_S$ and the dilution factor
$\xi_S$ for the model. In \cref{sec:Results} we 
will use this procedure to derive an independent 
set of dilution factors from our \textsc{tardis}
simulations.

\subsection{Dilution factors} \label{sec:DilutionTheory}
To understand the results of our numerical 
simulations, a firm grasp of
the basic physics behind the dilution factors is 
essential.  One of the most important effects in 
this context and the original motivation for the
introduction of the dilution factor
\citep[see][]{Hershkowitz1986a,Hershkowitz1986b} is 
the dilution of continuum
radiation that occurs in a scattering-dominated 
environment. If, as in SN~II, the scattering 
opacity greatly exceeds the absorptive opacity, a
thermally created photon can travel large optical 
depths before a true
absorption event returns it to the thermal energy 
pool. As a result, these photons can escape the 
ejecta without thermalizing and can efficiently 
carry away thermal energy from deep inside the 
atmosphere. This allows the intensity
of the radiation field to fall below the thermal 
value ($B_\mathrm{\nu}$) but to still 
resemble the spectral energy distribution of a blackbody. 
From random walk arguments, it can be shown 
\citep[see e.g.,][]{Mihalas1978} that
the relevant optical depth for this process, 
usually referred to as the thermalization depth 
$\tau_\mathrm{thm}$, scales roughly like
\begin{equation}
  \tau_\mathrm{thm} \propto 
  \sqrt{\frac{\chi_{\nu}}{3 \chi_{\nu, \mathrm{abs}}}}
\end{equation}
where $\chi_{\nu}$ denotes the total opacity and 
$\chi_{\nu,\mathrm{abs}}$ the
absorptive component. Under these conditions, the 
emergent flux resembles that of a blackbody with 
the temperature at the thermalization depth but is 
diluted by an amount 
$\xi^2  \approx 1 /\tau_\mathrm{thm}$.

\section{Example spectra} \label{sec:example_spectra}
\footnotetext[1]{\href{http://www.weizmann.ac.il/astrophysics/wiserep/}{
http://www.weizmann.ac.il/astrophysics/wiserep/}}
In \cref{sec:Method}, a detailed description of our 
efforts to extend \textsc{tardis} for the spectral 
modeling of \sniis{} has been given.  Here, we
apply the extended code to calculate synthetic 
spectra for two epochs of SN1999em, a prototypical 
event of this class. Our goal is to demonstrate 
that, with the implemented changes, we are able to 
reproduce the spectral properties
of such normal hydrogen-rich supernovae. Since we 
do not aim to perform a quantitative spectroscopic 
analysis, we have not extensively fine-tuned the
model to exactly fit the observations but have 
adopted parameters similar to
those used in previous studies by 
\citet{Baron2004} and \citet{Dessart2006}.
 
\begin{figure}
  \begin{center}
  \hspace*{-0.45cm}
  \vspace{-0.45cm}
    \includegraphics[width=0.45\textwidth]{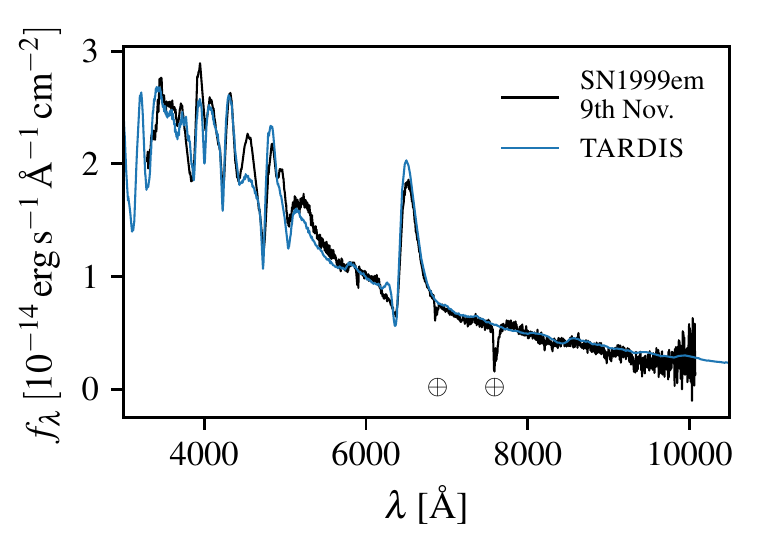}
  \end{center}
  \caption[]{\textsc{tardis} spectral model (blue) 
  for the observations of SN1999em (black) taken by 
  \citet{Hamuy2001} on the 9th of November. 
  We have smoothed the Monte Carlo spectrum using a
    Savitzky-Golay filter \citep{Savitzky1964}.
  The observational data
  has been taken from the
  \href{http://www.weizmann.ac.il/astrophysics/wiserep/}{WISeREP} archive
  \citep{Yaron2012}\footnotemark[1] and has been 
  de-reddened according to the 
  \citet*{Cardelli1989} law with a color excess of
  $E(B-V)=0.08$. To account for the peculiar 
  velocity of the host galaxy, the
  observations have been blueshifted by 
  \SI{770}{\km \per \second}
  \citep[see][]{Leonard2002}. Finally, the 
  synthetic spectrum has been scaled
  to match the observed de-reddened flux 
  $f_\lambda$. The main telluric features are 
  marked with circled crosses.
 }
  \label{fig:sn1999em_comp}
\end{figure}

As in \citet{Dessart2006}, we adopt a power-law 
density profile with index $n=10$ and
a CNO-enhanced composition with an otherwise solar 
metallicity for both epochs (see
\cref{sec:SNModel} for details). 
Our first model is for the 9th of November, 
corresponding to around two weeks
after explosion. At this point, the hydrogen 
envelope is still fully ionized
but the envelope has already cooled sufficiently 
for appreciable line blanketing by metals to 
develop. Apart from the very weak He \textsc{i}
5875\,\AA\,feature, helium lines have already 
disappeared from the spectrum.
Since the temperature is still too high for the Ca 
infrared triplet to form, the spectrum redwards of 
H$\alpha$ remains featureless. Our \textsc{tardis}
model nicely reproduces these characteristics as 
demonstrated by the comparison
to the observations taken by \citet{Hamuy2001} in 
\cref{fig:sn1999em_comp}. 
The observed spectrum has been de-reddened 
according to a color excess of
$E(B-V) = 0.08$, which is slightly less than the 
value of $E(B-V) = 0.1$ chosen
in previous studies by 
\citet{Baron2004} and \citet{Dessart2006}. We have 
blueshifted the observations by
\SI{770}{\km \per \second} 
\citep[see][]{Leonard2002} to correct for the
peculiar velocity of the host galaxy. The only major
shortcoming of our model is that it underproduces 
the strength of the Fe \textsc{ii}
lines at $\sim$4550\,\AA\, and $\sim$5140\,\AA\,. 
Since this epoch coincides
with the recombination from Fe \textsc{iii} to Fe 
\textsc{ii}, the predicted strengths of these
features are, however, very sensitive to small 
changes in the parameters and to
the adopted ionization treatment.

\begin{figure}
  \begin{center}
  \hspace*{-0.45cm}
  \vspace{-0.45cm}
    \includegraphics[width=0.45\textwidth]{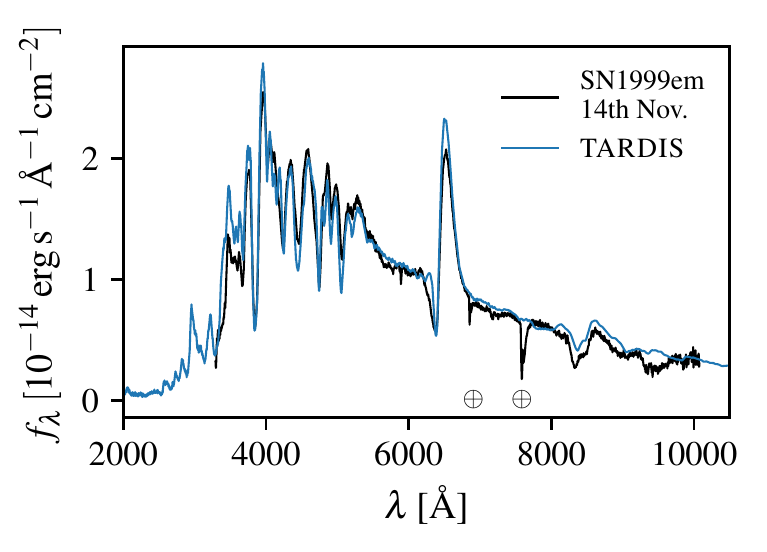}
  \end{center}
  \caption[]{Same as \cref{fig:sn1999em_comp} but 
  for the observations of
  SN1999em on the 14th of November.
  }
  \label{fig:sn1999em_1411_comp}
\end{figure}

The second epoch we are modeling corresponds to an 
intermediate stage in the photospheric-phase 
evolution of the supernova.
On the 14th of November, roughly 3 weeks after 
explosion, hydrogen recombination has set in and 
the spectrum shows very prominent
H$\alpha$ emission. Further cooling of the envelope 
has significantly strengthened the effect of line 
blanketing compared to the previous epoch. 
Redwards of H$\alpha$ the continuum is no longer 
featureless, since the temperature has dropped 
sufficiently for the Ca infrared triplet to appear.
\Cref{fig:sn1999em_1411_comp} shows a comparison of 
the spectrum taken by \citet{Hamuy2001} to our 
\textsc{tardis} model. 
We have corrected the observations for reddening 
and peculiar velocities in the
same fashion as for the first epoch. 
Overall, our synthetic spectrum reproduces the 
measured SED quite well. The two
prominent Ca features, Ca H\&K and the infrared 
triplet, are matched well in both strength and 
shape. However, our model slightly overestimates 
the H$\alpha$ emission, whereas the width of the 
absorption trough is underestimated. As mentioned 
by \citet{Dessart2006}, who found similar
problems, the latter might be related to blending 
with Fe \textsc{ii} and Si \textsc{ii} lines.

\section{Model grid} \label{sec:ModelGrid}
Having established the capabilities of the extended 
\textsc{tardis} version to produce accurate 
synthetic spectra for \sniis{}, we use the code to 
calculate an independent set of dilution factors. 
To this end, we have set up a grid of 343 models. 
These have been constructed to cover the 
interesting physical parameter space of 
inner boundary temperatures $T_\mathrm{inner}$ from 
\SIrange{9500}{24000}{\kelvin}, photospheric 
densities $\rho_\mathrm{ph}$ from 
\SIrange{7e-15}{8e-14}{g\per\cm\cubed}, power-law
density indexes $n$ from 6 to 14 and photospheric 
velocities $v_\mathrm{ph}$
from \SIrange{3000}{14000}{\km\per\second}. 
For these parameters the models span a range of effective 
temperatures $T_\mathrm{eff}$ from 
\SIrange{4900}{12000}{\kelvin}.
We take the photospheric properties 
$\rho_\mathrm{ph}$ and $v_\mathrm{ph}$ to refer to 
the position at which the electron scattering 
optical depth is $\tau = 2/3$. In practice, the 
models are set up by adopting $\rho_0 = 
\rho_\mathrm{*}$ and $r_0 = v_\mathrm{*} t$ for the 
density profile (\cref{eq:sn_density_profile}), 
where $\rho_\mathrm{*}$ and $v_\mathrm{*}$ are 
specific values selected from the desired range of 
photospheric density and velocity. To ensure that the photosphere of 
the model will lie at the appropriate depth, $t$ 
(time since explosion) is estimated using
\begin{equation} \label{eq:t_model}
t = \frac{2 (n - 1) m_H \mu_e}{3 v_\mathrm{*} \rho_\mathrm{*} \sigma_T} 
\end{equation}
where $m_H$ is the mass of a hydrogen atom, 
$\mu_\mathrm{e}$ is the mean
molecular weight per electron and $\sigma_T$ is the 
Thomson cross section. In making this estimate, it 
is assumed that $\mu_e$ is well-approximated by 
$\mu_e = 1.52$, as appropriate for a composition of 
ionized hydrogen and singly-ionized helium. 
The full calculation is then carried out and the 
true values of $\rho_\mathrm{ph}$ and 
$v_\mathrm{ph}$ are extracted from the simulation. 
In general, the true values of the photospheric 
parameters ($\rho_\mathrm{ph}$, $v_\mathrm{ph}$) 
are very close to the originally selected reference 
parameters ($\rho_\mathrm{*}$, $v_\mathrm{*}$) from 
which the model was generated. Nevertheless, we 
always refer to each model by the derived 
(simulation) values of $\rho_\mathrm{ph}$ and 
$v_\mathrm{ph}$.
We note that the true inner boundary of our 
computational domain lies considerably deeper than 
the (approximate) photosphere, typically at $\tau 
\sim 27$.

\begin{figure}
  \begin{center}
  \hspace*{-0.45cm}
    \includegraphics[]{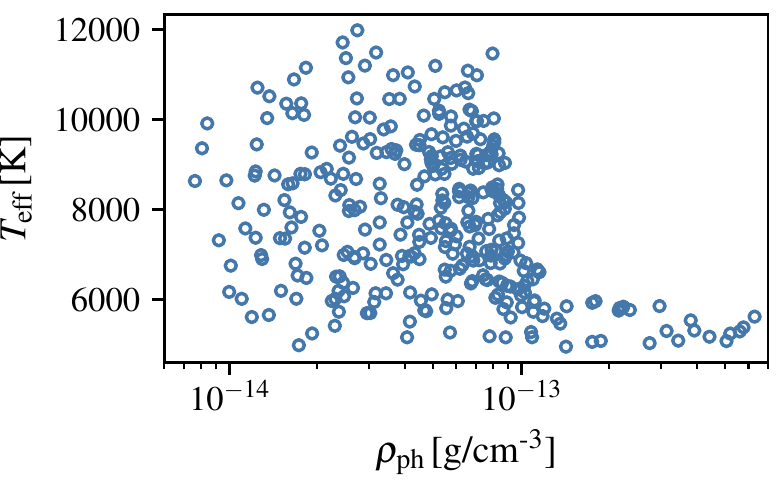}
  \end{center}
  \caption[]{Illustration of the model grid from 
  \cref{sec:ModelGrid}. The plot
  shows the effective temperature 
  $T_\mathrm{eff}$ and the photospheric
  density $\rho_\mathrm{ph}$ for all atmosphere 
  models. At low temperatures the actual 
  photospheric density can exceed the targeted 
  upper limit of \SI{8e-14}{g\per\cm\cubed}, due 
  the development of a strong recombination wave, 
  which complicates the mapping between 
  photospheric properties and the computational 
  grid.}
  \label{fig:Grid_T_ne}
\end{figure}

The setup of the model grid is done in the form of 
a latin hypercube design
\citep{Stein1987}. In this approach, each 
parameter range is subdivided
into $N$ equal intervals, where $N$ is the number 
of models and one random parameter value is 
selected from each subinterval. This guarantees 
that, in contrast to a conventional cartesian grid, 
$N$ distinct values exist for each
parameter. This is in particular beneficial if the 
quantities of interest are only weakly
sensitive to a subset of parameters. 
\footnote[2]{ Consider the extreme case that one or 
more parameters have no influence at all. For the 
hypercube each point still contains new 
information, whereas for the cartesian mesh most of 
the grid has become redundant.}
Finally, to illustrate the properties of our set of 
models, a projection of the
grid in the $T_{\mathrm{eff}}$ - 
$\rho_\mathrm{ph}$ plane is shown in
\cref{fig:Grid_T_ne}.  As can be seen, the desired 
parameter space is for the most part
uniformly covered.  Deviations from the uniform 
spacing arise from the use of \cref{eq:t_model} to 
map between photospheric quantities and the
computational grid. This process becomes less 
reliable as soon as a strong
recombination front develops.  
The main motivation for using a mostly uniform grid 
is that it allows us to study the differential 
influence of model parameters, such
as the photospheric density, on the dilution 
factors. In this context,
correlations between the input parameters have to 
be avoided as far as possible. However, since 
quantities such as photospheric temperature and
density are certainly not completely independent in 
nature, this also means that the grid includes 
models that are not representative of normal 
\sniis{}. One example would be an object with a 
very high expansion velocity but an
extremely low temperature.

\section{Results} \label{sec:Results}
\subsection{Overview}
\begin{figure}
  \begin{center}
  \hspace*{-0.45cm}
    \includegraphics[]{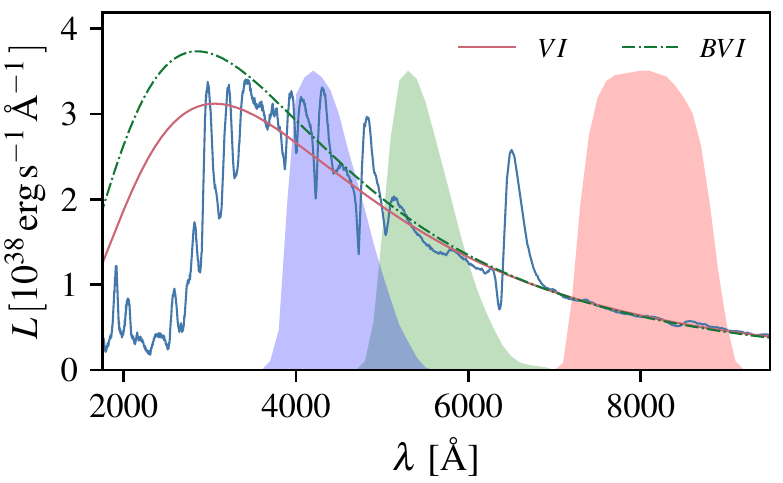}
  \end{center}
  \caption[]{Comparison of the dilute blackbody 
  models from \{\textit{V},\textit{I}\} (red) and 
  \{\textit{B},\textit{V},\textit{I}\}
    synthetic photometry (green) to the original 
    \tardis{} spectrum (blue). The
    spectrum is our SN1999em model for the 9th of 
    November (see \cref{fig:sn1999em_comp}). 
    The parameters of the blackbody fits are
    $T_{VI} = \SI{9500}{\kelvin}$, 
    $T_{BVI} = \SI{10200}{\kelvin}$ and
    $\xi_{VI} = 0.49$, $\xi_{BVI} = 0.45$. We have 
    overplotted the transmission
    curves for the \textit{B}, \textit{V} and 
    \textit{I} filters from
    \citet{Bessell1990}. The Monte Carlo spectrum 
    has been smoothed using a
    Savitzky-Golay filter \citep{Savitzky1964}.}
  \label{fig:DiluteBBFit}
\end{figure}

\begin{figure*}
  \begin{center}
  \hspace*{-0.45cm}
    \includegraphics[width=\textwidth]{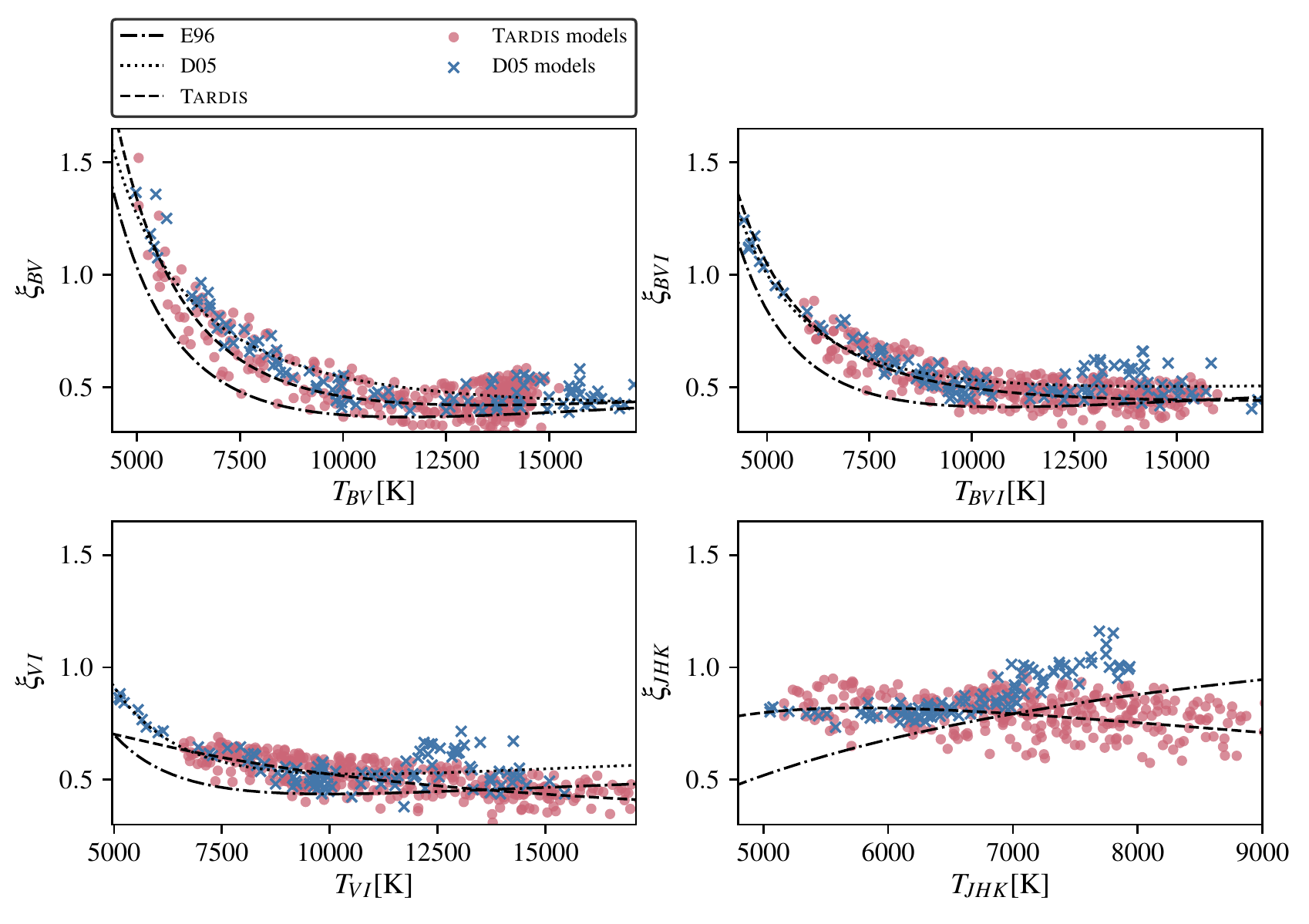}
  \end{center}
  \caption[]{Dilution factors $\xi_S$ as a function 
  of color temperature $T_S$
    for filter combinations 
    $S$=\{\textit{B},\textit{V}\},
    \{\textit{B},\textit{V},\textit{I}\}, 
    \{\textit{V},\textit{I}\} and
    \{\textit{J},\textit{H},\textit{K}\}. 
    We use a common y-axis scale for all bandpass 
    combinations $S$ to highlight the differences 
    in the scaling behavior of the dilution 
    factors. For comparison purposes, we include
    the models of \citetalias{Dessart2005a} as blue 
    crosses. Polynomial fits to the dilution 
    factors are shown for all sets of models 
    (dashed: \textsc{tardis}, dashed dotted: 
    \citetalias{Eastman1996}, dotted:
    \citetalias{Dessart2005a}). For 
    \{\textit{J},\textit{H},\textit{K}\} the
    \citetalias{Dessart2005a} curve is not included 
    due to a misprint in the tabulated fit 
    coefficients in the original paper.}
  \label{fig:DilutionOverview}
\end{figure*}

From synthetic photometry of our model spectra, we 
can derive color temperatures $T_S$ and dilution 
factors $\xi_S$ according to \Cref{eq:SynEpm}. To 
facilitate the comparison to the results of
\citetalias{Eastman1996} and 
\citetalias{Dessart2005a}, we focus our analysis
on the bandpass combinations 
$S$=\{\textit{B},\textit{V}\},
\{\textit{B},\textit{V},\textit{I}\}, 
\{\textit{V},\textit{I}\} and
\{\textit{J},\textit{H},\textit{K}\} with filter 
functions taken from
\citet{Bessell1988} and \citet{Bessell1990}. 
Examples of the dilute blackbody models constructed 
in the synthetic EPM analysis are shown in 
\cref{fig:DiluteBBFit}.
The results of our analysis are summarized in 
\cref{fig:DilutionOverview}, which
displays dilution factors $\xi_{S}$ and color 
temperatures $T_S$ for all our
models, as well as comparison values from 
\citetalias{Eastman1996} and
\citetalias{Dessart2005a}. The color temperatures 
constitute the most important parameters in the 
study of the dilution factors, since they account 
for most of the variance in the correction factors 
and can be easily inferred from observations.
Variations of the remaining parameters such as 
photospheric density or velocity are
in most cases of secondary importance and are 
responsible for the observed dispersion around the 
color-temperature trend. 
In \cref{fig:DilutionOverview} we find good 
agreement with \citetalias{Dessart2005a}, in 
particular at low to medium color temperatures.
For \{\textit{B},\textit{V},\textit{I}\} 
and \{\textit{V},\textit{I}\} the
results match well for temperatures below 
\SI{12500}{\kelvin} and for
\{\textit{J},\textit{H},\textit{K}\} for 
temperatures below \SI{7000}{\kelvin}.
In \{\textit{B},\textit{V}\} the dilution factors 
are similar to \citetalias{Dessart2005a} over the 
entire temperature range. For higher color
temperatures in 
\{\textit{B},\textit{V},\textit{I}\}, 
\{\textit{V},\textit{I}\}, 
and \{\textit{J},\textit{H},\textit{K}\} our models 
tend to be systematically
more dilute than \citetalias{Dessart2005a} with 
values closer to those published by 
\citetalias{Eastman1996}. As will be discussed in 
\cref{sec:DensityDiscussion}, part of this 
discrepancy can be attributed to differences in the 
adopted photospheric densities. We note that
for all bandpass combinations the intrinsic scatter 
of our dilution factors is
slightly larger than for the set of models by 
\citetalias{Dessart2005a}. This
was to be expected, since we have constructed our 
grid of models in such a way that at all 
temperatures the whole range of remaining 
parameters is covered (see \cref{sec:ModelGrid}).
In contrast, the dilution factors of 
\citetalias{Eastman1996} show a much
smaller dispersion, since only a narrow part of the 
parameter space is explored in their study. Before 
concluding our discussion, we stress that this 
scatter does not correspond to the diversity of 
real objects but only reflects our
ignorance about the parameter space occupied by 
SNe~II. Finally, following 
\citetalias{Eastman1996} and 
\citetalias{Dessart2005a} we present third order
polynomial fits to the color temperature dependence 
of our dilution factors
$\xi_S = \sum_i a_i (\SI{e4}{\kelvin}/T_S)^i$ in 
\cref{tab:FitCoeff}.

\begin{table}
  \centering
  \begin{tabular}{c|c|c|c|c}
     & \{\textit{B},\textit{V}\}  & \{\textit{B},\textit{V},\textit{I}\} & 
     \{\textit{V},\textit{I}\} & \{\textit{J},\textit{H},\textit{K}\}\\
    \hline\hline
    $a_0$ & 0.7417&  0.5356 & 0.2116 & -0.0384 \\
    $a_1$ & -0.8662 &  -0.3355 & 0.3799 & 0.9918 \\
    $a_2$ & 0.5828 & 0.2959 & -0.0673 &    -0.2867 \\
  \end{tabular}
  \caption[]{Coefficients $a_i$ of polynomial fits 
  $\xi_S = \sum_i a_i (\SI{e4}{\kelvin}/T_S)^i$ to 
  the models for bandpass combinations 
  $S$=\{\textit{B},\textit{V}\}, 
  \{\textit{B},\textit{V},\textit{I}\}, 
  \{\textit{V},\textit{I}\} and 
  \{\textit{J},\textit{H},\textit{K}\}.}.
  \label{tab:FitCoeff}
\end{table}

\subsection{Influence of atmospheric properties}
In the previous section, the dependence of the 
dilution factors on temperature
has been discussed. Here, we will focus on the 
effects of density. First, we analyze the variation 
of the dilution factors with photospheric density. 
Secondly, we investigate the influence of the 
steepness of the density profile. Finally, to 
conclude the discussion of the effect of model 
parameters, we will assess the robustness of the 
dilution factor fit curves against changes in the
metallicity.

\subsubsection*{Influence of the photopsheric density} 
\label{sec:EffectRhoPh}
\begin{figure}
  \begin{center}
  \hspace*{-0.45cm}
    \includegraphics[width=0.45\textwidth]{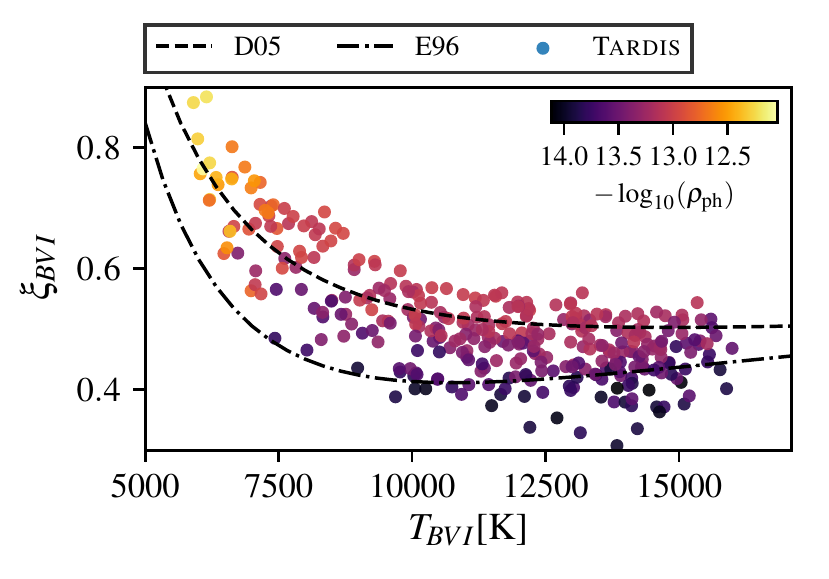}
  \end{center}
  \caption[]{Dilution factors $\xi_{BVI}$ as a 
  function of color temperature
  $T_{BVI}$. To illustrate the density dependence 
  of our \textsc{tardis} models, the logarithm of 
  the photospheric density
  $\log_{10}\rho_\mathrm{ph}$ is color-coded.  For 
  comparison purposes we
  include the polynomial fits to the dilution 
  factors of \citetalias{Eastman1996} 
  (dashed-dotted) and \citetalias{Dessart2005a}
(dashed).}
  \label{fig:xi_fit_nedependence}
\end{figure}

\begin{figure*}
  \begin{center}
  \hspace*{-0.45cm}
    \includegraphics[width=\textwidth]{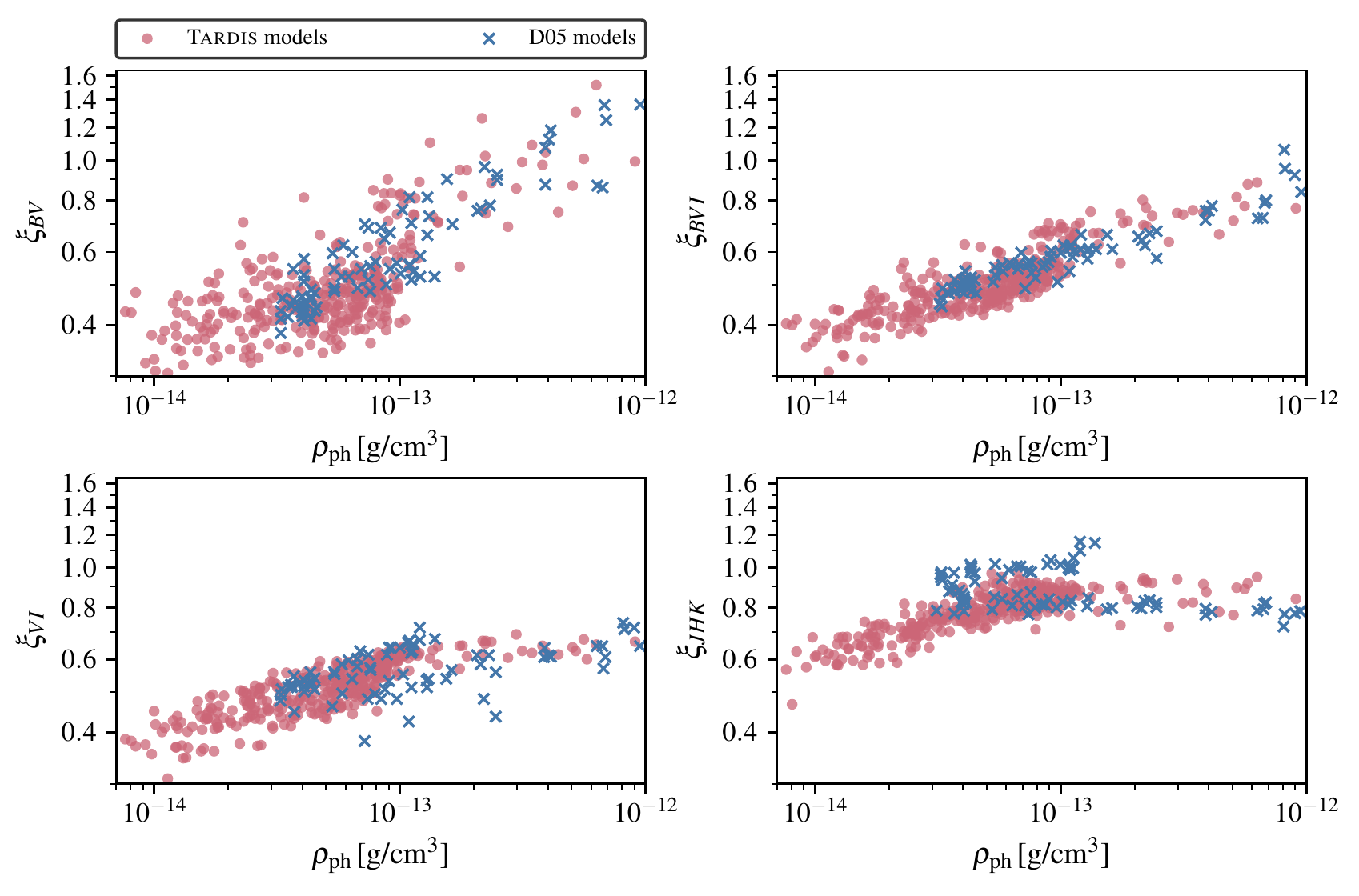}
  \end{center}
  \caption[]{Variation of the dilution factors 
  $\xi_S$ with photospheric density 
  $\rho_\mathrm{ph}$ for filter combinations 
  $S$=\{\textit{B},\textit{V}\},
    \{\textit{B},\textit{V},\textit{I}\}, 
    \{\textit{V},\textit{I}\} and
    \{\textit{J},\textit{H},\textit{K}\}. For 
    comparison purposes the models of 
    \citetalias{Dessart2005a} are shown as blue 
    crosses.\footnote{The information about the 
    \citetalias{Dessart2005a} shown have been 
    extracted from their paper. Note that not all 
    models presented by \citetalias{Dessart2005a} 
    are included in the illustration.}}
  \label{fig:DilutionOverviewRho}
\end{figure*}

Variations in the photospheric density account for 
most of the dispersion of the models around the 
general color temperature dependence as illustrated 
in \cref{fig:xi_fit_nedependence}.
Since the density cannot be easily constrained from 
observational data, it is essential to understand 
its influence on the dilution factors to quantify 
the associated uncertainties on the mean and the 
variance of the tabulated fit curves. 
\Cref{fig:DilutionOverviewRho} shows the density 
dependence of our dilution factors and a comparison 
to the results of \citetalias{Dessart2005a}. 
In general the dilution factors tend to increase 
with photospheric density, with the strength of the 
scaling varying between bandpass combinations. 
This behavior can be understood by remembering that 
the amount of continuum flux dilution depends on 
the ratio of continuum to scattering
opacity (see \cref{sec:DilutionTheory}). Since the 
main contribution to the
scattering opacity comes from Thomson scattering, 
it is proportional to the electron density. 
Thermalization processes, on the other hand, 
roughly scale with the square of the electron 
density.  As such, we expect the ratio of the two 
to increase with density, yielding a smaller amount 
of flux dilution at high
densities. To study this behavior in a more 
quantitative way, we adopt the same ansatz as 
\citetalias{Eastman1996} for the
dilution factors:
\begin{equation}
  \xi_S = \rho_\mathrm{ph}^{\gamma}z(T_S).
\end{equation}
Here, $z(T_S)$ denotes a polynomial of the same 
form as those used in
\cref{tab:FitCoeff}.  A least squares fit to our 
set of models yields 
the density scaling indexes $\gamma$ listed in 
\cref{tab:DensityScalingCoeff}.
Overall, the inferred density dependence of our 
dilution factors is moderate
with similar magnitudes for all passbands.
The scaling indexes $\gamma$ are systematically a 
bit larger than those published in 
\citetalias{Eastman1996} 
with the largest difference in the infrared  (see
\cref{tab:FitCoeff}). However, as illustrated by 
\cref{fig:DilutionOverviewRho} the density 
dependence in the infrared is not well described by 
a single power-law for the entire density range.
In \cref{sec:DensityDiscussion} we will use the 
inferred power-law scalings to assess whether 
differences in the assumed photospheric densities 
play an import role in understanding the 
discrepancies between the published sets of 
dilution factors.
\begin{table}
  \centering
  \begin{tabular}{c|p{12mm}|p{10mm}|p{10mm}|p{10mm}|p{10mm}}
     &\centering\arraybackslash Models& \centering\arraybackslash 
     \{\textit{B},\textit{V}\}  & \centering\arraybackslash 
     \{\textit{B},\textit{V},\textit{I}\} & \centering\arraybackslash
     \{\textit{V},\textit{I}\} &\centering\arraybackslash 
     \{\textit{J},\textit{H},\textit{K}\}\\
\hline
$\gamma$& \centering\arraybackslash \textsc{tardis} 
\newline \citetalias{Eastman1996} & \centering\arraybackslash0.106
\newline 0.0776&\centering\arraybackslash0.137 
\newline0.0933 &\centering\arraybackslash 0.133 
\newline 0.0769 &\centering\arraybackslash  0.116
\newline0.0307  \\
  \end{tabular}
  \caption[]{Coefficients $\gamma$ of polynomial 
  fits $\xi_S = \rho_\mathrm{ph}^{\gamma}z(T_S)$ 
  to the density dependence of the dilution
    factors for bandpass combinations 
    $S$=\{\textit{B},\textit{V}\},
    \{\textit{B},\textit{V},\textit{I}\}, 
    \{\textit{V},\textit{I}\} and
    \{\textit{J},\textit{H},\textit{K}\}. 
    Here, $z(s) = \sum_i a_i
    (\SI{e4}{\kelvin}/T_S)^i$ denotes a 
    polynomial of the same form as used in
    \cref{tab:FitCoeff} but with different fitted 
    values for the coefficients.}
  \label{tab:DensityScalingCoeff}
\end{table}

\subsubsection*{Influence of the density structure}
\begin{figure}
  \begin{center}
  \hspace*{-0.45cm}
    \includegraphics[]{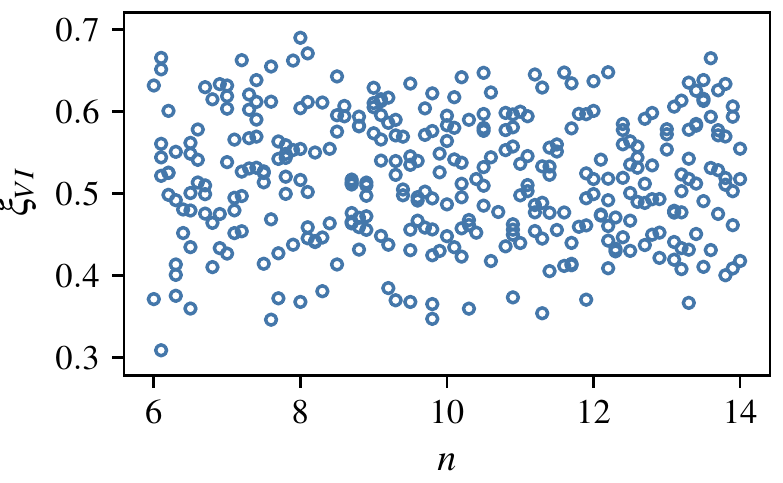}
  \end{center}
  \caption[]{Dilution factors $\xi$ for the 
  bandpass combination \{\textit{V},\textit{I}\} 
  as a function of the density indexes
    $n=-\dif{\ln{\rho}}/\dif{\ln{r}}$ of the 
    power-law model atmospheres.} 
  \label{fig:NDependence}
\end{figure}

For a power-law atmosphere the ratio of 
photospheric density and density index
$n$ is approximately given by
\begin{equation}
  \frac{\rho_\mathrm{ph}}{n - 1} \approx 
  \frac{2}{3}\frac{\mu_\mathrm{e}}{\sigma_\mathrm{e} v_\mathrm{ph} t} \; \; .
\end{equation} 
For a given outflow ionization, time of
explosion and expansion velocity, an increase in 
the density index $n$ results in higher 
photospheric densities $\rho_\mathrm{ph}$ and 
therefore less flux dilution. However, if the 
density structure is treated as an independent 
parameter, the dilution factors do not show a 
strong dependence on $n$ as illustrated by 
\cref{fig:NDependence}. For a possible explanation, 
we refer the reader to \citetalias{Eastman1996}, 
who have found the same behavior and have proposed 
a physical motivation in their \S 3.3.

\subsubsection*{Influence of metallicity}
\begin{figure}
  \begin{center}
  \hspace*{-0.45cm}
    \includegraphics[]{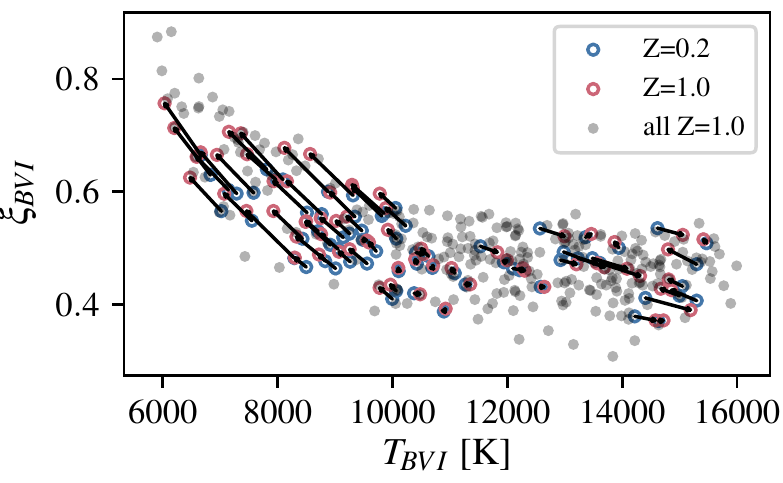}
  \end{center}
  \caption[]{Change of dilution factors $\xi_{BVI}$ 
  with metallicity $Z$. 
  The subset of the original solar metallicty 
  models for which corresponding
  calculations with a lower metallicity have been 
  performed are shown in red.
  The metal-poor models are depicted in blue.  
  Arrows illustrate the changes in
  model properties induced by the change to the 
  subsolar metallicity of
$Z=0.2$. To facilitate the comparison to the 
general $\xi$-$T$ trend, all
models of the original grid are included in gray.}
  \label{fig:xis_scaled}
\end{figure}

Line blanketing by metals, in particular iron group 
elements, plays an important role in shaping the 
spectral energy distribution of \sniis{} and the
resulting influence of metallicity on the emergent 
spectrum has been discussed in detail in the 
literature \citep[see e.g.,][]{Dessart2005}. 
However, neither
\citetalias{Eastman1996} nor 
\citetalias{Dessart2005a} discuss in depth how
this effect translates into changes in the dilution 
factors. To investigate the sensitivity of the 
$\xi$-$T$ fit curves to changes in metallicity, 
we have rerun a random subset of 68 models of our 
solar metallicity grid ($Z=1$) with a lower 
metallicity of $Z=0.2$. The resulting changes in 
the color temperatures and dilution factors are 
shown in \cref{fig:xis_scaled} for the 
\{\textit{B},\textit{V},\textit{I}\} bandpass
combination. As expected, at high temperatures 
($T_{BVI} \gtrapprox \SI{10000}{\kelvin}$) 
the influence of metallicity on the model 
properties is negligible, since the degree of 
ionization is too high for significant
line blanketing to develop in the optical and near-UV.
\footnote{The seemingly random 
displacements for models at high color temperatures 
are an artifact resulting from the flattening of 
the color-color temperature relationship. As a 
consequence small changes in the fluxes can induce 
large changes in the inferred temperatures.} For 
moderate and low temperatures large changes in the 
color temperature up to thousands of degrees are 
observed. However, the associated
changes in the dilution factors are approximately 
aligned with the general scaling of $\xi$ with $T$. 
Compared to the intrinsic scatter of the models, 
the induced changes in the functional behavior of 
$\xi$ with $T$ are of secondary importance. 
Thus, in the investigated regime, ranging from 
solar to distinctly subsolar, the tabulated fit 
curves are robust against modifications
of the metallicity. 

\section{Comparison to previous studies}\label{sec:prev_studies}
\subsection{Radiative transfer}
To put our results into context, we review 
differences in the radiative
transfer modeling between the \textsc{Cmfgen} 
code used by \citetalias{Dessart2005a},
the \textsc{Eddington} code used by \citetalias{Eastman1996} and 
\textsc{tardis} and discuss
possible effects on the dilution factors.  The main 
differences lie in the ionization treatment of 
metal species, the handling of line opacity and the
inclusion of relativistic transfer effects.  

For the calculations presented by
\citetalias{Dessart2005a}, only the effect of the 
Doppler shift on the frequency of the radiation 
field is taken into account.
\citetalias{Eastman1996} follow a different 
approach based on the premise that
radiation-field time dependence can be included in 
a quasi-static treatment by enforcing a constant 
luminosity in the comoving frame. In this case, the
time-dependent comoving-frame transport equation 
reduces to a much simpler expression that
differs from \citetalias{Dessart2005a} only by an 
additional term $\beta I_\nu / r$, where $I_\nu$ is 
the specific intensity of the radiation field. 
This term is formally identical to the part of the 
full transport equation that describes the redshift 
of photons in the scattering
process and thus the adiabatic loss of radiation 
energy.  However, the sign is
changed and the magnitude decreased by a factor of three. 
Both approaches neglect the so-called advection 
term that arises from the frame
transformation of angles \citep[see e.g.,][for a 
discussion]{Pistinner1994}. This term is generally 
deemed to be more important than the aberration 
term \citep[see e.g.,][]{Baron1996a}. Taking into 
account the additional reduction of
the magnitude of the aberration term in 
\citetalias{Eastman1996}, we conclude
that the differences in handling the relativistic 
terms between \citetalias{Eastman1996} and 
\citetalias{Dessart2005a} are small in comparison
to our relativistic treatment 
(\cref{sec:RelTrans}), which corresponds to a
full solution of the quasi-static relativistic 
transport problem. Since we can
achieve good agreement with 
\citetalias{Dessart2005a} despite this difference, 
we consider it unlikely that
relativistic effects play an important role in 
explaining the systematic offset
between \citetalias{Eastman1996} and 
\citetalias{Dessart2005a}.

Another possible source for discrepancies, which 
has been discussed previously
in the literature \citepalias[see][]{Dessart2005a}, 
is the treatment of line interactions. Here, the 
differences start with the handling of the opacity.
Both \textsc{Cmfgen} \citepalias{Dessart2005a} and 
\textsc{tardis} treat the
contributions of all lines to the opacity 
individually, in a consistent manner.
In contrast, \textsc{Eddington} 
\citepalias{Eastman1996} adopts the more
convenient but approximate expansion opacity 
formalism of \citet{Eastman1993}
that combines all line opacity in a wavelength bin. 
For the opacity calculation, the
expansion opacity formalism in 
\citetalias{Eastman1996}, as well as the method
used in \textsc{tardis}, rely on the Sobolev 
approximation \citep{Sobolev1957},
whereas \textsc{Cmfgen} adopts the comoving-frame 
method. For micro-turbulent
velocities of less than $\SI{100}{\km \per 
\second}$, as adopted in
\citetalias{Dessart2005a}, the Sobolev method is of 
similar accuracy as the comoving-frame method in 
describing the formation of hydrogen lines in 
SNe~II \citep[see][]{Duschinger1995f}. In regions 
where line overlap is possible, in
particular in the metal line forest in the blue, 
the Sobolev approximation may,
however, be less accurate than the comoving-frame 
method. 

With respect to line
interactions, the final difference between the 
codes concerns the redistribution of the absorbed 
radiation. Only \textsc{Cmfgen} computes a full
NLTE source function for all included species.  In 
\citetalias{Eastman1996}
line interactions are treated in detail only for a 
few selected elements, in most cases only
hydrogen. For the remaining species, resonance 
scattering is assumed and
effects such as fluorescence or collisional deexcitations are neglected.
\textsc{tardis} strikes a balance between the 
approximate treatment of
\citetalias{Eastman1996} and the full NLTE 
calculation of
\citetalias{Dessart2005a}. In principle, our 
implementation of the macro atom
scheme of \citet{Lucy2002,Lucy2003} also provides a 
full NLTE description of
the redistribution process. However, in our 
simulations only radiative and
collisional bound-bound transitions are included 
for species other than hydrogen. Despite this 
simplification, the approximate NLTE emissivities 
from the macro atom 
provide a full treatment of fluorescence. It is 
also worth pointing out that
the predicted emissivities are largely insensitive 
to errors in the excited states population and 
therefore to our use of approximate excitation 
treatments. This has been demonstrated by 
\citet{Lucy2002} and may be understood by
considering that, in the context of the macro atom 
machinery, the most relevant
level number densities are those of the ground 
state and low-lying metastable
levels. Radiative excitations from these states 
account for most of the activations of
the macro atom and their populations are likely to 
be close to LTE with respect to the ground state. 
In contrast, the level number 
densities of excited states, which will be less
accurately estimated, are not as important in 
setting the rate of macro atom
activations and enter in the emissivity only 
through minor modifications of the
internal redistribution probabilities for 
stimulated emission. Thus we argue that
the macro atom approach captures most of the 
essential physics of a full NLTE
treatment as opposed to the resonance scattering 
approximation used in \citetalias{Eastman1996}. 
As such, it constitutes a promising source of
systematic discrepancies between 
\citetalias{Eastman1996} on the one hand and
\citetalias{Dessart2005a} and \textsc{tardis} on 
the other hand.

To conclude our discussion of the major differences 
in the numerical treatments, we compare the 
different methods used for calculating the
ionization state of metal species. An accurate 
solution to the ionization
balance is essential in modeling the line 
blanketing, which shapes the
spectral energy distribution in the blue. Due to 
the use of super-levels,
\textsc{Cmfgen} \citepalias{Dessart2005a} is able 
to consistently, though approximately, treat all
species in NLTE. In contrast, 
\citetalias{Eastman1996} calculated the ionization
using NLTE only for a few selected species and only 
for a subset of their atmospheric models. For the 
remaining models and species, LTE at the
electron temperature is assumed. 
Similarly, \textsc{tardis} relies on simplified 
prescriptions for the calculation of the ionization 
balance of metals. For the results presented in
this paper, we have used the nebular ionization approximation of 
\citet{Mazzali1993}. In principle, this 
method should provide a more accurate description 
of the ionization balance in a diluted, radiation-
dominated environment than the assumption of LTE. 
However, neither assumption can fully replace a 
detailed photoionization calculation. Still, our 
spectral models for SN1999em (see
\cref{sec:example_spectra}) reproduce the observed 
line blanketing well. This instills confidence 
that, at least for the early and intermediate stage
evolution, the nebular ionization treatment 
adequately captures the essential physics. 

Ultimately, it is extremely difficult to assess the 
extent to which, if at all,
individual numerical differences contribute to the 
systematic discrepancy between the sets of dilution 
factors.  Based on qualitative arguments, we have
deemed it unlikely that the handling of 
relativistic terms plays an important
role in this context. 
We have identified the use of the
very simple resonance scattering approximation by 
\citetalias{Eastman1996} as one of the main 
distinguishing features from both our and 
\citetalias{Dessart2005a}'s numerical approaches. 
As such, it can be regarded as a promising possible 
contributory factor to the systematic differences. 
However, these interpretations are speculative and 
should be taken with a grain of salt.

\subsection{Effect of model grid assumptions} 
\label{sec:DensityDiscussion}
In the previous section we have discussed how 
differences in the radiative transfer calculations 
can affect the dilution factors. In the context
of the discrepancy between \citetalias{Eastman1996} 
and \citetalias{Dessart2005a} most of the 
discussion in the literature \citep[see
e.g.,][]{Dessart2005a,Jones2009} has revolved around 
these issues. However, another possibly important 
(albeit banal) source of systematic differences is 
the choice of model grid properties. To demonstrate 
this, we have modified the plot depicting the 
temperature dependence of our dilution factors
for the \{\textit{B},\textit{V},\textit{I}\} 
bandpass combination to include
the color-coded photospheric density for each model 
(see \cref{fig:xi_fit_nedependence}). From this, it 
is obvious that the inferred fit
curves can easily be moved upwards or downwards by 
preferentially sampling either the high density or 
the low density regions of the parameter space.
Since the exact distribution and correlation of 
parameters such as density,
temperature and velocity are not known for the 
population of SNe~II, there
exists a certain amount of freedom in the setup of 
the model grid.  

To quantitatively illustrate the role such effects 
may have, we have investigated the influence of 
density in particular. For a comparative study we 
need to consider families of models for which color 
temperature, densities and dilution factors are 
available -- accordingly, we make use of our 
\textsc{tardis} models, the 
\citetalias{Eastman1996} models and the models 
presented for the tailored EPM analysis
of SN1999em, SN2005s and SN2006bp 
\citep{Dessart2006,Dessart2008}. This set of
38 models covers the relevant range of color 
temperatures and generally follows
the fit curves published in 
\citetalias{Dessart2005a}.\footnote{We use the 
tailored EPM models because the full set of model 
parameter data has not been published for the 
calculations presented in 
\citetalias{Dessart2005a}.}

\begin{figure}
  \begin{center}
  \hspace*{-0.45cm}
  \vspace{-0.45cm}
    \includegraphics[width=0.45\textwidth]{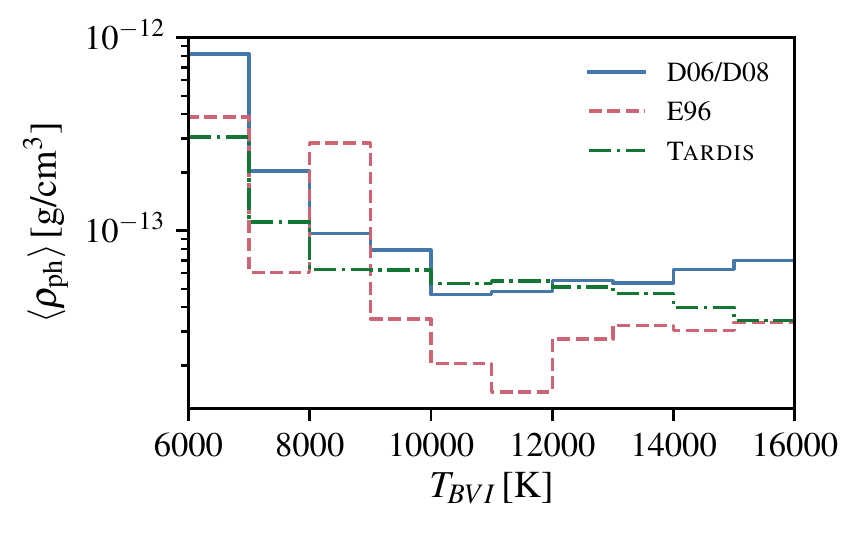}
  \end{center}
  \caption[]{Comparison of the mean photospheric 
  density $\langle \rho_\mathrm{ph} \rangle$ 
  at a given \{\textit{B},\textit{V},\textit{I}\}
    color temperature for the models from 
    \citetalias{Eastman1996} (red
    dashed), the tailored EPM analyses of 
    \citet{Dessart2006} and \citet{Dessart2008} 
    (D06/D08, blue solid), and this paper (green 
    dash-dotted). The plotted densities
    have been rescaled slightly to account for 
    differences in the composition as
    outlined in \cref{sec:DensityDiscussion}.}
  \label{fig:density_comp}
\end{figure}

\begin{figure}
\centering
\begin{subfigure}[]{0.45 \textwidth}
    \includegraphics[]{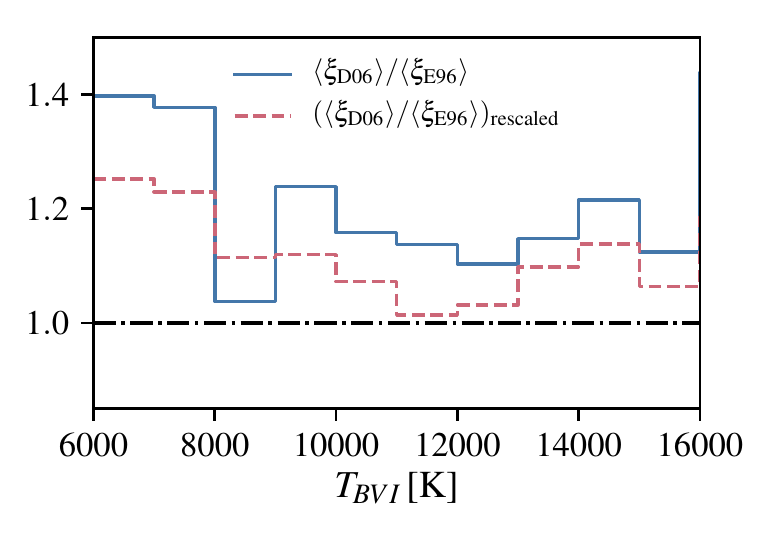}
    \caption{}
   \label{fig:E96_D06_rescale}
\end{subfigure} \\
\begin{subfigure}[]{0.45 \textwidth}
    \includegraphics[]{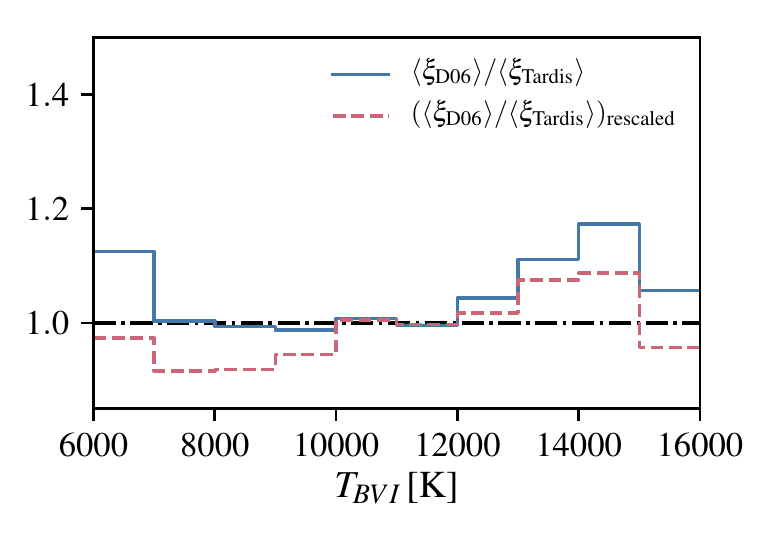}
    \caption{}
   \label{fig:Tardis_D06_rescale}
\end{subfigure} \\[2ex]
  \caption[]{Comparison of the discrepancy between 
  the dilution factor set in
    the \{\textit{B},\textit{V},\textit{I}\} 
    bandpass combination of
    \citet{Dessart2006} and \citet{Dessart2008} 
    $\langle \xi_\mathrm{D06} \rangle$, and
    \citetalias{Eastman1996} $\langle 
    \xi_\mathrm{E96} \rangle$ (upper panel,
    a), and with respect to the results of this 
    paper $\langle \xi_\mathrm{Tardis} \rangle$ 
    (lower panel, b). The findings before (blue
    solid) and after (red dashed) the application 
    of a density correction
    factor are shown. The details of the procedure 
    are described in
    \cref{sec:DensityDiscussion}.} 
  \label{fig:DensityDiscrepancy}
\end{figure}

Before we compare densities, we approximately 
correct for changes of the
electron densities between the set of models due to 
differences in composition (specifically, we 
rescale the densities from 
\citetalias{Eastman1996}'s 
models with the estimated ratio of the mean 
molecular weights per electron). 
The mean rescaled densities
$\langle \rho_\mathrm{ph} \rangle$ are shown in 
\cref{fig:density_comp} as a
function of \{\textit{B},\textit{V},\textit{I}\} 
color temperature. Overall, the densities used in 
this paper, and in \citet{Dessart2006} and 
\citet{Dessart2008} tend to be larger
than those of \citetalias{Eastman1996} with maximum
differences of a factor of a few.  The 
conspicuous jump in density for the 
\citetalias{Eastman1996} models between 8000 and 
9000\,K stems from two
exponential atmospheres (e12.2,e12.3). 
To check whether this density mismatch might 
alleviate some of the tension
between the dilution factors by 
\citetalias{Eastman1996}, and those of 
\citet{Dessart2005a,Dessart2006} and 
\citet{Dessart2008}, we rescale the dilution 
factors $\xi_\mathrm{S}$ using
the simple power-law relation $\xi_S \propto 
\rho_\mathrm{ph}^\gamma$ from
\cref{sec:EffectRhoPh}. For this purpose we do not 
use the mean densities
$\langle \rho_\mathrm{ph} \rangle$ from 
\cref{fig:DensityDiscrepancy} but the
appropriate average $\langle 
\rho_\mathrm{ph}^\gamma \rangle^{1/\gamma}$.
\cref{fig:E96_D06_rescale} illustrates the effect 
of the rescaling on the discrepancy between the two 
sets of models. Applying the density correction
reduces the maximum difference from roughly 40\% to 
20\%, but fails to remove
the systematic offset completely.  
As can be seen in \cref{fig:Tardis_D06_rescale}, 
the procedure is more successful for our set of 
dilution factors. After rescaling, only a maximum 
difference of around 8\% remains between our 
calculations and those of
\cite{Dessart2006} and \citet{Dessart2008}.  
We stress that due to the simplifying
assumptions we have made the results above are only 
qualitative in nature.
Nevertheless, our discussion demonstrates that 
differences in the setup of the
model grid, for example different choices for the 
photospheric densities, can
introduce systematic uncertainties on the 10\% 
level in the dilution factors.
This most likely explains part of the discrepancy 
between the results of \citetalias{Eastman1996}, 
and those of \citet{Dessart2005a,Dessart2006} and 
\citet{Dessart2008}. To eliminate this 
additional error source, approaches are needed that
strongly constrain the relevant parameters through 
observational data.  One possibility would be to 
base the dilution factor fit curves on the tailored 
EPM analyses of a representative set of \sniisp{}.

\section{Conclusions} \label{sec:Conclusions}
In this work, we present an extension of the 
Monte Carlo radiative transfer code \textsc{tardis} 
to the spectral synthesis of \sniisp{}. The key 
feature of our numerical approach is an updated
radiation--matter interaction scheme, which 
provides a full treatment of bound-bound, 
bound-free, free-free and collisional processes 
based on the macro atom scheme of 
\citet{Lucy2002,Lucy2003}. 
The second major improvement concerns the 
calculation of the plasma state. The
code now contains a self-consistent determination 
of the thermal structure from the heating and 
cooling balance as well as a full NLTE calculation 
of the ionization and excitation state for 
hydrogen. Other changes include an
improved handling of relativistic effects, an 
adaption of the spectral synthesis calculation for 
high optical depths and a different initialization 
of the plasma state. We demonstrate the 
capabilities of the extended code by
modeling two different epochs of the prototypical 
SN~II SN1999em. For both epochs good agreement with 
the observed spectra is achieved, instilling
confidence that \textsc{tardis} is well-suited for 
quantitative spectroscopic analysis of 
photospheric-phase, hydrogen-rich supernovae.

In line with our goal to use \textsc{tardis} for 
measuring distances, our final
application is the calculation of an independent 
set of EPM dilution factors.
In this context, a long-standing issue has been the 
systematic discrepancy of around 20\% between the 
results of \citetalias{Eastman1996} and
\citetalias{Dessart2005a}, which translates into an 
uncertainty of the EPM
distance of the same magnitude. To address this 
problem, we have performed
radiative transfer calculations for a set of 343 
\textsc{tardis} models, which
span a wide range of temperatures, densities and 
expansion velocities.  Despite
using significantly different numerical techniques, 
the dilution factors extracted
from these calculations show good agreement with 
those published by \citetalias{Dessart2005a}. This 
result helps remove some of the tension
between the available sets of distance correction 
factors. It is still somewhat
unclear which differences in the numerical approach 
make the models of
\citetalias{Eastman1996} systematically more dilute 
than ours and \citetalias{Dessart2005a}'s. 
Based on our calculations, we can plausibly rule 
out only one of the previously
suggested explanations, namely differences in the 
treatment of relativistic effects. 

Our other focus lay on investigating the parameter 
dependences of the dilution factors. Similar to 
\citetalias{Eastman1996} and 
\citetalias{Dessart2005a}, we identify density as 
one of the most important parameters in setting the 
magnitudes of the dilution factors. 
Our power-law fits to the density dependence yield 
similar scaling behaviors as for the calculations 
by \citetalias{Eastman1996}. As in
\citetalias{Eastman1996}, we do not find a strong 
effect of the steepness of the density profile.
In addition, we have demonstrated that changing the 
metallicity from solar to decidedly subsolar 
(Z=0.2) only induces minor modifications in the 
relationship between color temperature and dilution factors.

Finally, we have investigated differences in the 
setup of the model grid as an
additional source of systematic errors. In our 
discussion, we have demonstrated
that part of the discrepancy between 
\citetalias{Eastman1996} and
\citetalias{Dessart2005a} can plausibly be tracked 
back to differences in the assumed photospheric 
densities. This result highlights the need to base
tabulated dilution factors on approaches that 
constrain the model parameters
and their correlations more strongly through 
observational data. One way to
achieve this would be to apply the tailored EPM technique 
\citep{Dessart2006,Dessart2008} to a representative 
set of SNe~II.

In this paper we have established \textsc{tardis} 
as a new independent numerical tool for modeling 
\sniisp{} and have demonstrated its capability to
calculate accurate dilution factors. As a next step 
we plan to apply the code to measure absolute 
distances using the tailored EPM method 
\citep{Dessart2006,Dessart2008} or SEAM
\citep{Baron1995,Baron1996,Baron2004,Baron2007}. 
As a consequence of the inclusion of a much more 
detailed treatment of the radiative 
transfer process, the typical runtime 
of the \textsc{tardis} spectral synthesis procedure 
has increased from minutes needed in the original 
implementation by \citet{Kerzendorf2014} to hours. 
However, in light of the ubiquity of machine 
learning techniques and the continuous increase in 
computational resources, this increase in 
computational complexity is of minor concern and it 
will, for the first time, be feasible to perform 
the spectral fitting process in an automated 
manner. In combination with sampling techniques the
parameter space can be explored in a systematic 
manner and uncertainties in the
estimated parameters can be obtained. This will put 
strong constraints on the
accuracy of absolute distance measurements of 
\sniis{} and will help to assess their promise as 
tools for cosmology.

\begin{acknowledgements}
CV thanks Andreas Floers, Stefan Lietzau and Markus 
Kromer for stimulating discussions during
various stages of this project. The authors 
gratefully acknowledge Stefan
Taubenberger for sharing his enormous expertise in 
the field of supernovae and
for being a key member of the supernova cosmology 
project that motivated this work.
CV would also like to thank Michael Klauser for 
providing help and guidance during
the start of this project. The authors thank the anonymous reviewer 
for valuable comments.
This work has been supported by the Transregional 
Collaborative Research Center
TRR33 ``The Dark Universe'' of the Deutsche 
Forschungsgemeinschaft and by the 
Cluster of Excellence ``Origin and Structure of the 
Universe'' at Munich Technical
University. SAS acknowledges support from STFC 
through grant, ST/P000312/1. 
Data analysis and visualization was performed using 
\textsc{Matplotlib} \citep{Hunter2007},
\textsc{Numpy} \citep{numpy} and \textsc{SciPy} 
\citep{Scipy}.
\end{acknowledgements}

\bibliographystyle{aa} 
\bibliography{Mendeley,tardis,references}
\end{document}